\documentclass{emulateapj}

\usepackage{aas_macros}
\usepackage{url}
\usepackage{graphicx}
\usepackage{epsfig}
\usepackage{epstopdf}
\usepackage{amsmath}
\usepackage{hyperref}
\hypersetup{
    bookmarks=true,         % show bookmarks bar?
    unicode=false,          % non-Latin characters in AcrobatÕs bookmarks
    pdftoolbar=true,        % show AcrobatÕs toolbar?
    pdfmenubar=true,        % show AcrobatÕs menu?
    pdffitwindow=false,     % window fit to page when opened
    pdftitle={},    % title
    pdfauthor={Dr. Aaron Rizzuto},     % author
    pdfsubject={},   % subject of the document
    pdfcreator={Dr. Aaron Rizzuto},   % creator of the document
    pdfproducer={},  % producer of the document
    pdfkeywords={} {} {}, % list of keywords
    pdfnewwindow=true,      % links in new window
    colorlinks=false,       % false: boxed links; true: colored links
    linkcolor=red,          % color of internal links
    citecolor=green,        % color of links to bibliography
    filecolor=magenta,      % color of file links
    urlcolor=cyan           % color of external links
}
\newcommand{\Msun}{M$_{\odot}$~}

\begin{document}

\title{Dynamical Masses of Young Stars I: Discordant Model Ages of Upper Scorpius}

\author{
Aaron C. Rizzuto\altaffilmark{1},
Michael J. Ireland\altaffilmark{2},
Trent J. Dupuy\altaffilmark{1},
Adam L. Kraus\altaffilmark{1}
}

\altaffiltext{1}{Department of Astronomy, The University of Texas at Austin, Austin, TX 78712, USA}
\altaffiltext{2}{Research School of Astronomy \& Astrophysics, Australian National University, Canberra, ACT 2611, Australia}

\begin{abstract}
We present the results of a long term orbit monitoring program, using sparse aperture masking observations taken with NIRC2 on the Keck-II telescope, of seven G to M-type members of the Upper Scorpius subgroup of the Sco-Cen OB association. We present astrometry and derived orbital elements of the binary systems we have monitored, and also determine the age, component masses, distance and reddening for each system using the orbital solutions and multi-band photometry, including Hubble Space Telescope photometry, and a Bayesian fitting procedure. We find that the models can be forced into agreement with any individual system by assuming an age, but that age
is not consistent across the mass range of our sample. The G-type binary systems in our sample have model ages of $\sim$11.5\,Myr, which is consistent with the latest age estimates for Upper Scorpius, while the M-type binary systems have significantly younger model ages of $\sim$7\,Myr. Based on our fits, this age discrepancy in the models corresponds to a luminosity under-prediction of 0.8-0.15\,dex, or equivalently an effective temperature over-prediction of 100-300\,K for M-type stars at a given premain-sequence age. We also find that the M-type binary system RXJ1550.0-2312 has an age ($\sim$16\,Myr) and distance ($\sim$90\,pc)  indicating that it is either a nearby young binary system or a member of the Upper-Centaurus-Lupus subgroup with a 57\% probability of membership.
\end{abstract}


%%TARGET STAR INFORMATION TABLE (\label{targets})
\begin{deluxetable*}{ccccccccc}
\tabletypesize{\footnotesize}
\tablewidth{0pt}
\tablecaption{Keck-II/NIRC2 Orbit Monitoring Sample}
\tablehead{
\colhead{2MASS} & \colhead{Name} & \colhead{R.A.} & \colhead{Decl.} & \colhead{SpT} & \colhead{Ref} & \colhead{r$'$}&\colhead{K}&\colhead{W$_3$}
\\
& &\colhead{(J2000)} & \colhead{(J2000)} &&&\colhead{(mag)}&\colhead{(mag)}&\colhead{(mag)}
}
\startdata

J16081474-1908327 & GSC 6209-735                   & 16 08 14.744 & -19 08 32.77  & K2 & (1)  &11.01 &  8.426  & 8.226 \\%P98-44
J16245136-2239325 & GSC 6794-156                   & 16 24 51.363 & -22 39 32.54  &G6 & (1)   & 9.50 &  7.084   &  6.983 \\%P98-75
J16051791-2024195 & USco J160517.9-202420 & 16 05 17.915 & -20 24 19.65  &M3 & (2)  & 13.50 &  9.143 &  8.771 \\%P02-065
J15573430-2321123 & ScoPMS 17			      & 15 57 34.305 & -23 21 12.32  & M1 & (3) & 12.50$^a$ & 8.992  &  8.296 \\%ScoPMS17
J15500499-2311537 & RX J1550.0-2312                & 15 50 04.992 & -23 11 53.73  &M2	& (4)	&  13.34 & 8.930 &8.533  \\%K99-46
J16015822-2008121 & RX J1601.9-2008	      & 16 01 58.225 & -20 08 12.15  &G5	& (4) & 10.09 & 7.672  & 7.521 \\%K99-80
J16321179-2440213 & ROXs 47A                            &16 32 11.794 & -24 40 21.37   &M3	& (5)	& 12.84   & 7.929 & 6.509  \\%ROXs 47A

\enddata
\tablecomments{Basic information for the stars in our orbit monitoring sample. The references for the spectral types are: (1) Spectral types are taken from \citet{preibisch98}, (2) \citet{preibisch02}, (3) \citet{walter94}, (4) \citet{kunkel2000}, (5)  \citet{mcclure2010}.  The r$'$ magnitudes are taken from the APASS catalog, except for ScoPMS 17 ($^a$) which is taken from USNO-A2 in the absence of an APASS measurement, and the K magnitudes are taken from 2MASS. The final column lists the WISE band 3 (12\,$\mu$m) magnitudes \citep{wise10}. Note that with the exception of ROXs 47A, the sole Ophiuchus binary in our sample, these stars do not show evidence for the presence of a circumstellar disk \citep{luhman2012_disk}}
\label{targets}
\end{deluxetable*}

\section{Introduction}
\label{intro}
The majority of the Galactic star formation is likely to occur in embedded clusters containing massive stars \citep{lada03}. Such embedded clusters dissipate and dissolve, leaving unbound OB associations. Stars within these OB associations, and in particular their subgroups, are thought to be a coeval population that share a common star formation history, chemical abundance, and velocity \citep{zeeuw99}. 

 These young OB associations provide a glimpse into the state of a group of stars directly after formation. The nearest OB association to the sun is the Scorpius-Centaurus-Lupus-Crux association (Sco-Cen), and is also our closest location of recent high-mass star formation ($\sim$150\,pc). The association contains $\sim$150 B-type stars which are spatially grouped into three sub-groups: Upper-Scorpius, Upper-Centaurus-Lupus (UCL) and Lower-Centaurus-Crux (LCC), and provides one of the richest nearby laboratories for the study of star and exoplanet formation \citep{myfirstpaper}.
The coeval Sco-Cen populations are often used as ``age-calibrated" samples of objects in the study of a number of different science goals, such as circumstellar disk evolution \citep{carpenter09, chen11,rizzuto12}, exoplanet identification and evolution \citep{lafreniere08,lafreniere09,mikegsc6214} and multiplicity studies \citep{kouwenhoven07,kraus11,scopaper2}. Furthermore, mass estimation from models of very low mass companions to K and M-type association members is highly dependent on the assumed age. Thus, it is critical that the age estimation for young associations is both highly accurate, and unbiased. 

The age of the Sco-Cen subgroups is contentious. The youngest and most compact subgroup, Upper-Scorpius, was first age-dated using the main sequence turn-off, and was estimated to be $\sim$5-7\,Myr old \citep{geus92}. This age was supported by a spectroscopic survey for low-mass association members by \citet{preibisch02}, which also determined an population age of $\sim$5\,Myr according to the latest models of the time,  with a very narrow age spread. There is now some evidence for a spread in HR-diagram apparent age which correlates with Li abundance \citep{wifes1_2015}, and recent work utilizing new spectral typing and photometry of F-type members, and re-analysis of B/A-type members, has shown that Upper-Scorpius may have median age of  $\sim$11\,Myr, which is significantly different from any previous work \citep{pecaut12}. 

There is still some uncertainty in the ages of the various B-type members of Upper Scorpius:  While there are B-type Upper Scorpius stars which appear to have ages consistent with the $\sim$11\,Myr age estimation, there are others, such as $\tau$ Sco and $\omega$ Sco which are clearly very young. In the case of  $\tau$ Sco this is verified independently as it has  a well-determined temperature of 32000$\pm$1000\,K, luminosity of $\log{\mathrm{L}/\mathrm{L}_{\odot}} = 4.47\pm0.13$, and mass or 11$\pm$4\,\Msun measured from the He and H absorption lines \citep{simondiaz06}. Combined with the very slow rotation period of 43\,days  \citep{simondiaz06,strassmeier09}, this means that $\tau$ Sco is certainly very young, with an age  between 2 and 5\,Myr depending on the choice of models. For non-rotating stellar models  the most massive 15\,Myr stars should be approximately 12\,M$_{\odot}$ \citep{ekstrom12}. The existence of stars like $\tau$ Sco and $\omega$ Sco cloud the clear picture of single-epoch star formation in the Upper Scorpius region, and perhaps argue for some star formation between the age of $\rho$ Ophiuchus and the new canonical $\sim$11\,Myr age of Upper Scorpius.

The older subgroups of Sco-Cen also have somewhat unclear ages. The B, A and F-type UCL and LCC members have main-sequence turn off/on ages of  $\sim16 - 18$\,Myr \citep{mamajek02}, while studies of the incomplete sample of Lithium-rich G, K and M-type members show a variety of mass-dependent age estimates. The HR-diagram age for the known K-type stars in UCL and LCC is $\sim$12\,Myr, the few known M-type stars indicate a significantly younger age of $\sim$4\,Myr, and the G-type members have an age of 17\,Myr which is consistent with the more massive stars \citep{preibisch08, song12}. There is also a positional trend in the age of the PMS stars of the older subgroups, with stars closer to the Galactic Plane appearing significantly younger than objects further north. This is almost certainly the result of as yet undiscovered and un-clarified substructure within the older subgroups, which have a very complex star-formation history or unclear selection biases.

To date, the best large-scale age estimation for the Sco-Cen association and its subgroups has been purely photometric, meaning that colors and magnitudes are used to place association members on a theoretical HR diagram, and then fit model isochrones. This method suffers from a number of issues beyond the inescapable model dependancy. Firstly, unknown binarity effects observed photometry: An equal mass binary will be 0.7 magnitudes brighter than the either component, but will not change in color. In addition, fitting isochrones to photometry is highly dependent on the distance measure to the object being used. In particular, it has been shown that {\it HIPPARCOS} parallax measurements can potentially be incorrect for high-mass binary systems \citep{susilamsco}. This is further complicated by the effect of the significant rotation of some B-type stars on HR-diagram positioning. 

Comparison of HR-diagram positions of young stars to model isochrones also suffers from both the uncertainty intrinsic to the different evolutionary models, and the uncertainty in the spectral-type to effective temperature conversion for PMS stars.  The various evolutionary models can predict masses differing by up to 50\% for solar and sub-solar PMS stars \citep{hillenbrand04,stassun14}, and the uncertainty in the spectral-type to effective temperature conversion for young stars can be as large as 100-200\,K \citep{luhman03}.

A currently underutilized improvement for both age-dating Sco-Cen and evaluating the accuracy of PMS stellar evolution models at young ages is the inclusion of orbital mass information for well characterized Sco-Cen member binary system (e.g. \citealt{simon13,schaefer14,kraususcoctio5}). An accurate orbit can provide a direct measurement of the  total system dynamical mass, which can then be used as an additional, orthogonal dimension in model fitting. Paired with one or more contrast ratios between the primary and the secondary companion at different wavelengths, this provides a vast improvement in estimating the age of individual stars, which will then provide a indications of the age of the accompanying Upper Scorpius population. 

In this paper, we present the orbits  for seven low-mass Upper Scorpius stars monitored with Sparse Aperture Masking (SAM) techniques with the NIRC2 Camera on the Keck-II telescope. Using these seven orbits we estimate stellar parameters for these binary systems, including age, component masses and parallax, using  a Bayesian model fitting algorithm. We then present some evaluation of the available PMS models for predicting mass and luminosity of young stars of different temperatures, and provide an orbital estimate for the age of the Upper Scorpius subgroup of Sco-Cen.

%%\section{Target Sample, Observations, and Data Analysis}
\section{Target Sample}
We present a sample consisting of six Upper Scorpius binary systems (two G-type, one K-type and three M-type) and a single Ophiuchus region M-type binary system. These seven systems are the first completed orbits of a larger and ongoing orbit monitoring program of Sco-Cen, Taurus and Ophiuchus binary members. Table \ref{targets} lists basic stellar properties for the seven binary systems. The six Upper Scorpius members in this study were selected from a multiplicity and planet-search survey of the Upper-Scorpius subgroup of Sco-Cen \citep{bd1}, the targets for which were compiled in \citet{kraus07} from the wider literature of membership surveys of the Upper-Scorpius region \citep{preibisch98,preibisch01,preibisch02,slesnick06,ardila2000,walter94,martin04}. The final object, ROXs 47A, a hierarchical triple system first identified by \citet{barsony03}, is a member of the nearby $\rho$-Ophiuchus  star forming region which overlaps the Upper Scorpius region of sky. Following the identification of these systems as binary stars,  all of these targets were flagged as ideal for orbit monitoring due to their small angular separations and projected periods which were on the order of a few years.

\section{Keck NIRC2 Observations and Analysis}
We have monitored the seven systems in our sample using NIRC2 aperture masking in natural guide star AO mode, on a yearly basis, over a time-scale of approximately five years. All NIRC2 AO observations were taken using the smallest available pixel scale of 9.95\,mas per pixel \citep{yelda10}, and either a two or four location dither pattern. The majority of the observations were taken with the 9-hole aperture mask and the narrow-band CH$_4$S filter, with some observations in J,K$'$ and L$'$. 
%%NIRC2 KECK MASKING ORBIT POINTS TABLE %%%%%
\begin{deluxetable*}{cccccc}
\tabletypesize{\footnotesize}
\tablewidth{0.6\textwidth}
\tablecaption{Keck-II/NIRC2 Astrometry}
\tablehead{
 \colhead{Date} & \colhead{MJD}& \colhead{Filter}& \colhead{$\rho$}& \colhead{$\theta$}& \colhead{$\Delta m$}\\
&&&\colhead{(mas)}& \colhead{($^\circ$)}& \colhead{(mag)}
}
\startdata
\multicolumn{6}{c}{GSC6209-735}\\
\hline
23/06/15 & 57196   & CH4S  & 26.0$\pm$3.1 & 219.7$\pm$2.0 & 3.81$\pm$0.18 \\
30/07/14 & 56868   & CH4S   &17.2$\pm$1.7  &175.8$\pm$6.4 &3.05 (Fixed)\\
7/08/13   & 56511   & CH4S  & 16.4$\pm$2.5 & 72.5$\pm$1.2  & 2.90$\pm$0.50 \\
4/04/12   & 56021   & CH4S & 33.1$\pm$0.6 & 31.8$\pm$0.7& 3.09$\pm$0.03 \\
22/06/11 & 55734   & CH4S & 26.8$\pm$1.0 & 11.6$\pm$0.9 & 2.99$\pm$0.05 \\
5/04/10  & 55291    & CH4S & 12.7$\pm$1.0 & 243.5$\pm$5.8 & 3.10$\pm$0.01 \\
1/06/09  & 54983    & CH4S & 26.1$\pm$1.6 & 195.5$\pm$1.1 & 3.50$\pm$0.10 \\
30/05/07 & 54250   & CH4S & 31.0$\pm$2.0 & 42.5$\pm$3.6 & 3.15$\pm$0.01 \\
\hline
\multicolumn{6}{c}{GSC6794-156}\\
\hline
23/06/15 & 57196   & CH4S & 63.0$\pm$0.1 & 58.9$\pm$0.2 & 0.51$\pm$0.01\\
6/08/13 & 56510     & Kc & 70.5$\pm$0.1 & 93.5$\pm$0.1 & 0.46$\pm$0.01 \\
5/06/11 & 55717     & L' & 70.5$\pm$0.1 & 129.8$\pm$0.1 & 0.45$\pm$0.01 \\

4/04/10 & 55290    & Jc & 66.0$\pm$0.1 & 150.1$\pm$0.1 & 0.53$\pm$0.01 \\

1/06/09 & 54983    & CH4S & 60.6$\pm$ 0.1 & 167.5$\pm$0.1 & 0.50$\pm$0.01 \\

6/06/07 & 54257    & K' & 44.3$\pm$0.1 & 230.7$\pm$0.1 & 0.45$\pm$0.01 \\
\hline
\multicolumn{6}{c}{USco J160517.9-202420}\\
\hline
30/07/14& 56868 & CH4S& 25.1$\pm$0.2&303.8$\pm$0.6          &0.39$\pm$0.02\\
4/04/12 & 56021  & CH4S & 31.73$\pm$0.04 & 271.19$\pm$0.1 & 0.44$\pm$0.01 \\

22/06/11 & 55734  & CH4S & 37.20$\pm$0.08 & 278.6$\pm$0.1 & 0.46$\pm$0.01 \\

5/04/10 & 55291    & CH4S & 38.28$\pm$0.07 & 287.6$\pm$0.1 & 0.39$\pm$0.01 \\

1/06/09 & 54983     & CH4S & 33.6$\pm$0.1 & 294.3$\pm$0.2 & 0.41$\pm$0.01 \\

17/06/08 & 54634    & CH4S & 21.4$\pm$0.1 & 309.2$\pm$0.5 & 0.52$\pm$0.02 \\

6/06/07 & 54257    & K' & 16.2$\pm$0.6 & 251.1$\pm$1.1 & 0.4$\pm$0.07 \\
\hline
\multicolumn{6}{c}{ScoPMS 17}\\
\hline
29/08/14 & 56867	& Kc & 34.4$\pm$0.1&	117.8$\pm$0.1&	0.71$\pm$0.01\\
7/08/13 & 56511    & CH4S & 27.10$\pm$0.04 & 132.5$\pm$0.1 & 0.79$\pm$0.01 \\

4/04/12 & 56021    & CH4S & 11.7$\pm$0.3 & 198.1$\pm$3.5 & 0.79$\pm$0.01 \\

22/06/11 & 55734    & CH4S & 22.5$\pm$0.10 & 32.7$\pm$0.2 & 0.79$\pm$0.02 \\

5/04/10 & 55291   & CH4S & 39.7$\pm$0.10 & 50.5$\pm$0.1 & 0.76$\pm$0.01 \\

5/06/07 & 54256   & K' & 53.9$\pm$0.2 & 68.9$\pm$0.2 & 0.78$\pm$0.01 \\
\hline
\multicolumn{6}{c}{RXJ1550.0-2312}\\
\hline
30/07/14 & 56868	&CH4S & 42.3$\pm$2.2& 136.7$\pm$3.0&1.2$\pm$0.2\\
4/04/12 & 56021    & CH4S & 66.96$\pm$0.20 & 89.0$\pm$0.1 & 0.86$\pm$0.01 \\
22/06/11 & 55734  & CH4S & 66.80$\pm$0.30 & 77.5$\pm$0.3 & 0.87$\pm$0.03 \\

5/04/10 & 55291    & CH4S & 58.66$\pm$0.15 & 56.7$\pm$0.1 & 0.89$\pm$0.01 \\

1/06/09 & 54983    & CH4S & 46.33$\pm$0.12 & 36.0$\pm$0.1 & 0.82$\pm$0.01 \\

17/06/08 & 54634    & CH4S & 26.88$\pm$0.17 & 344.0$\pm$0.3 & 0.81$\pm$0.01 \\

6/06/07 & 54257     & K' & 26.93$\pm$0.04 & 222.1$\pm$0.1 & 0.76$\pm$0.01 \\

5/06/07 & 54256    & K' & 26.95$\pm$0.05 & 222.1$\pm$0.1 & 0.76$\pm$0.01 \\
\hline
\multicolumn{6}{c}{RXJ1601.9-2008}\\
\hline
23/06/15 & 57196 & CH4S & 38.0$\pm$0.5 & 215.1$\pm$0.7 & 2.0$\pm$0.04\\
29/07/14 &  56867& K'  & 32.8$\pm$0.4&206.1$\pm$0.3&1.84$\pm$0.02\\
4/04/12 & 56021   & CH4S & 17.5$\pm$1.7 & 100.5$\pm$1.2 & 2.3$\pm$0.3 \\

22/06/11 & 55734   & CH4S & 25.5$\pm$0.3 & 67.6$\pm$0.3 & 2.04$\pm$0.02 \\

5/04/10 & 55291   & CH4S & 28.5$\pm$0.4 & 43.8$\pm$0.3 & 2.08$\pm$0.02 \\

17/06/08 & 54634     & CH4S & 24.4$\pm$0.7 & 231.5$\pm$0.8 & 2.08$\pm$0.05 \\

31/05/07 & 54251   & CH4S & 39.3$\pm$1.6 & 217.7$\pm$0.6 & 2.1$\pm$0.1 \\
\hline
\multicolumn{6}{c}{ROXs 47A}\\
\hline
29/07/14 & 56867 &  K'	& 52.3$\pm$0.2	& 73.1$\pm$0.2&0.09$\pm$0.01 \\
7/08/13 & 56511   & CH4S & 42.39$\pm$0.04 & 62.2$\pm$0.1 & 0.37$\pm$0.01 \\
22/06/11 & 55734 & J & 21.2$\pm$0.3 & 141.1$\pm$1.6 & 0.21$\pm$0.02 \\
5/04/10 & 55291 & CH4S & 43.4$\pm$0.2 & 108.5$\pm$0.2 & 0.22$\pm$0.01 \\
1/06/09 & 54983 & CH4S & 51.8$\pm$0.2 & 98.7$\pm$0.2 & 0.17$\pm$0.01 \\
24/05/02$^a$ & 52418 & K' & 40$\pm$30 & 107$\pm$20 & 0.1$\pm$1.60 \\
\enddata
\tablecomments{The full list of Keck NIRC2 observations of our low-mass aperture masking sample in the Upper Scorpius subgroup of Sco-Cen. The data provided are: angular separation $(\rho)$, uncertainty on separation $(\sigma_{\rho})$, companion position angle $(\theta)$, position angle uncertainty $(\sigma_{\theta})$, magnitude difference $(\Delta m)$, and magnitude difference uncertainty $(\sigma_{\Delta m})$. The original observation of ROXs 47a (labelled $^a$), was taken from the discovery paper of \citet{barsony03}, which used the Hale 200\,inch telescope, this measurement was not included in the orbital fit.}
\label{orbital_points_nrm}
\end{deluxetable*}

Aperture masking data reduction utilizes the complex triple product or closure-phase, in addition to squared  interferometric visibilities, in order to remove non-common path errors and variable optical aberrations. Binary system profiles can then be fit to the visibilities and closure phases to produce separations and position angles. A full explanation of the reduction and closure-phase fitting procedure is given in the appendix of \citet{bd1}. In Table \ref{orbital_points_nrm} we list the individual aperture masking observation details, including observation filter, fitted separations and position angle.

\section{Orbit Fitting}

\begin{figure}
\includegraphics[width=0.5\textwidth]{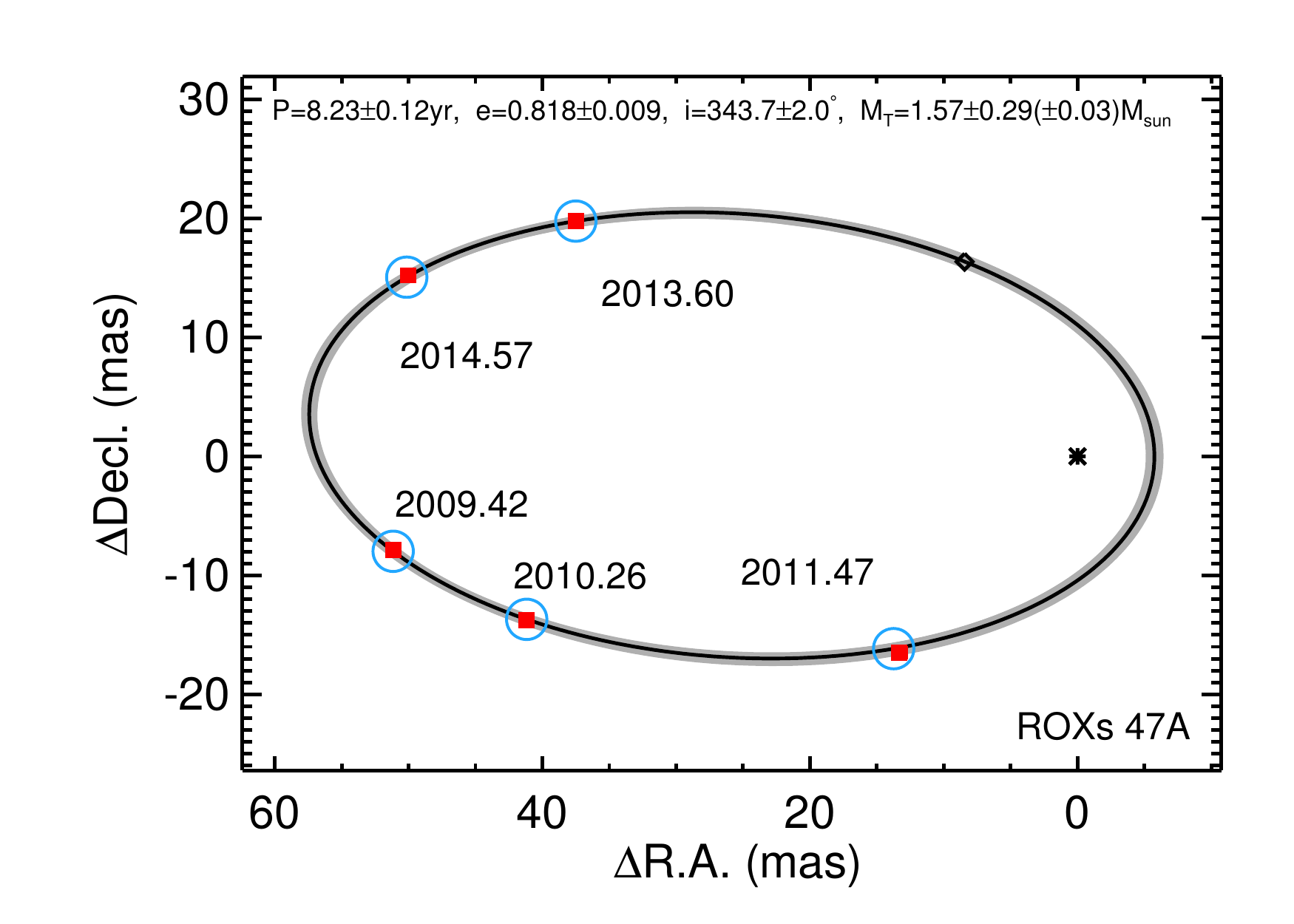}
\caption{Orbital solution for Ophiuchus binary ROXs 47Aab from NIRC2 AO aperture masking. The data is shown as red squares, with corresponding model fits shown as blue circles. We show the best fit orbit in black, with a 1-$\sigma$ region shaded in grey. The black diamond shows the orbital position of the secondary at the time of the corresponding HST observations (see Table \ref{hstphot_delta}). We also display the orbital period, semimajor axis, eccentricity and system mass at 145\,pc. The two uncertainties listed for the mass are derived by either including or excluding the $\pm$15\,pc distance uncertainty associated with Upper Scorpius membership. The bracketed uncertain, which only includes the uncertainty on the orbital parameters is representative of the system mass precision attainable with future GAIA parallaxes.}
\label{roxs47a_orbfit}
\end{figure}

%%%%%%%%%%%%%%%%%%%%%%%%%%%%%%%%%%%%%%%%
%%ORBITAL FITS FIGURES%%%%
\begin{figure*}
\includegraphics[width=0.5\textwidth]{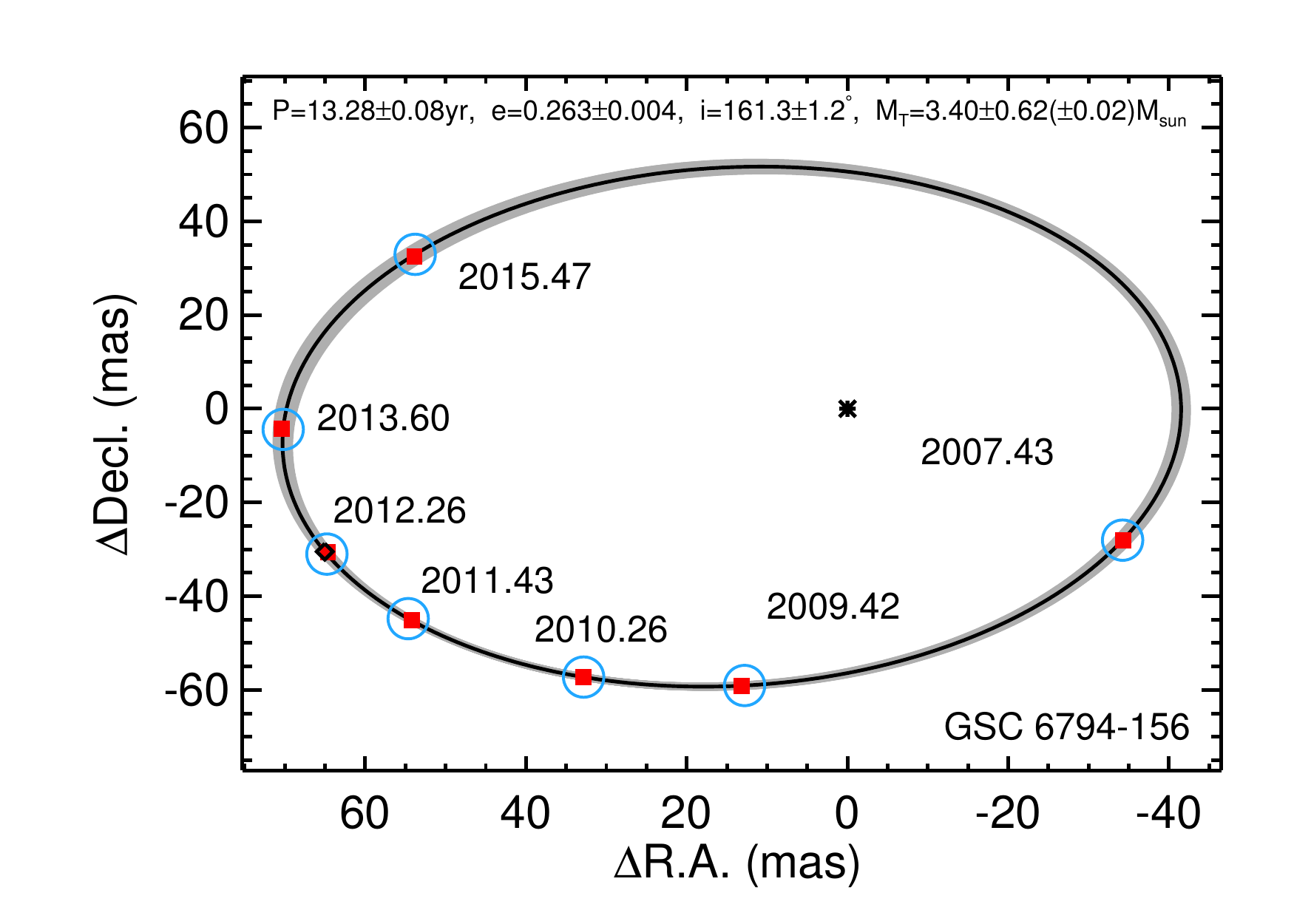}
\includegraphics[width=0.5\textwidth]{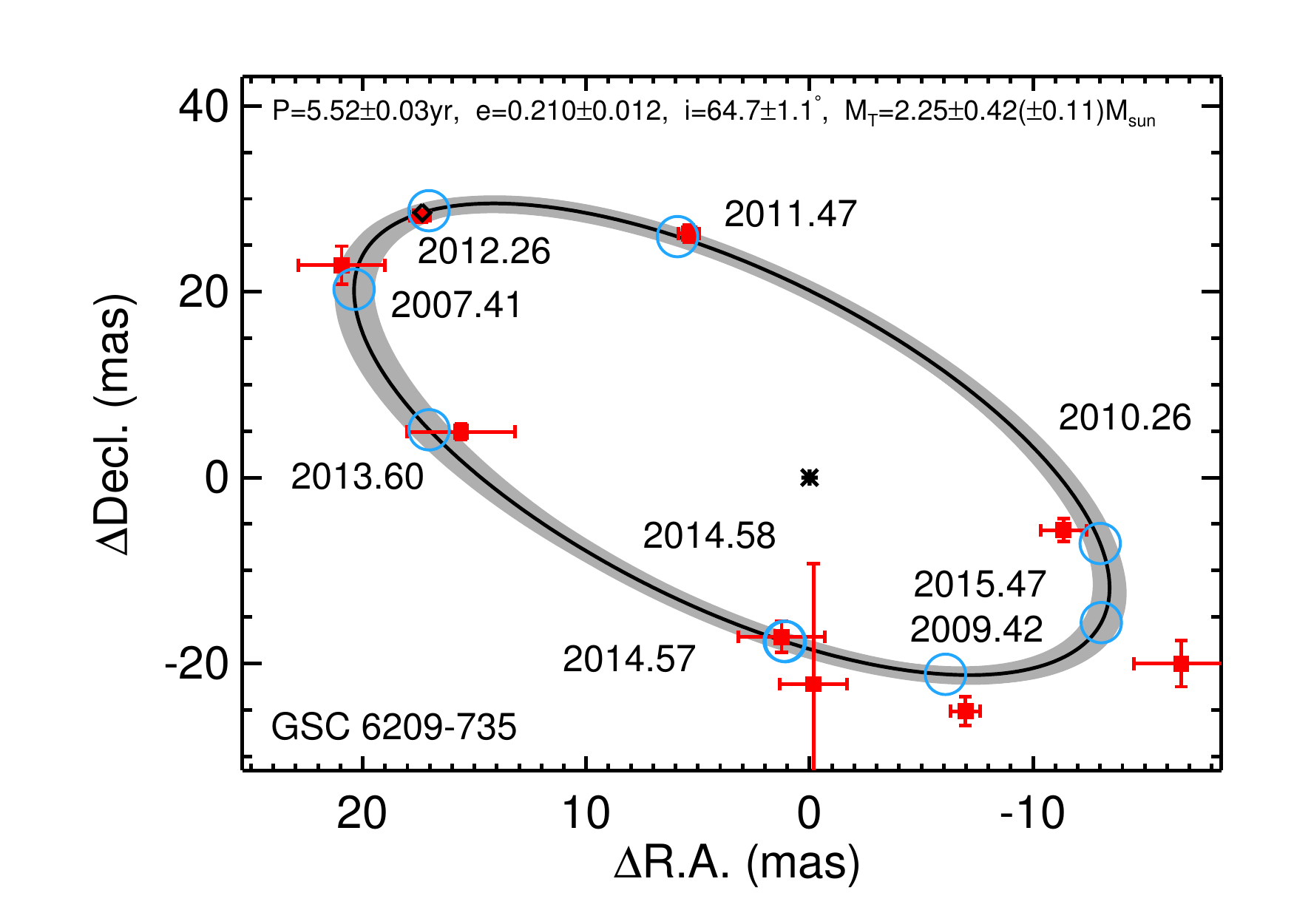}\\
\includegraphics[width=0.5\textwidth]{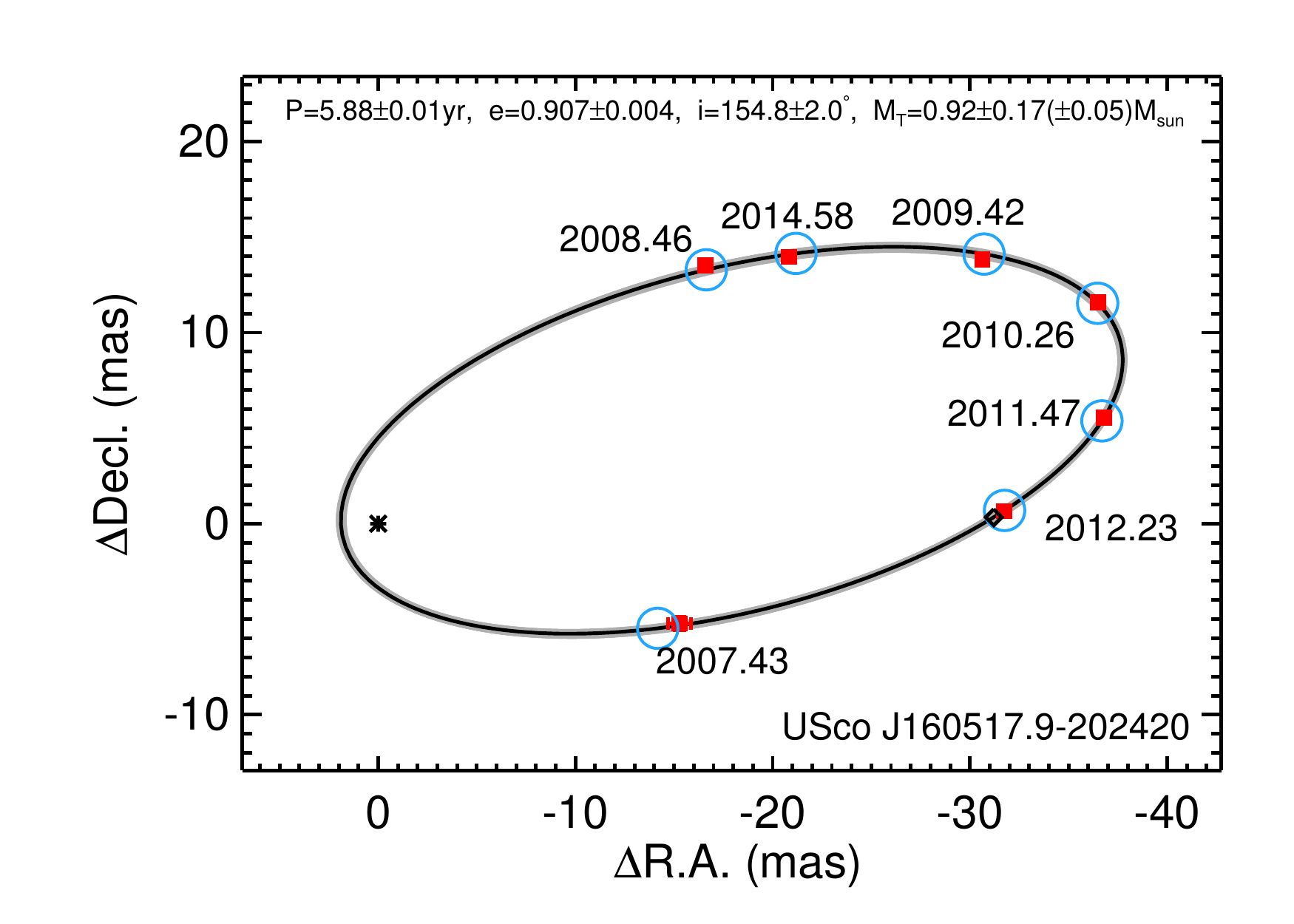}
\includegraphics[width=0.5\textwidth]{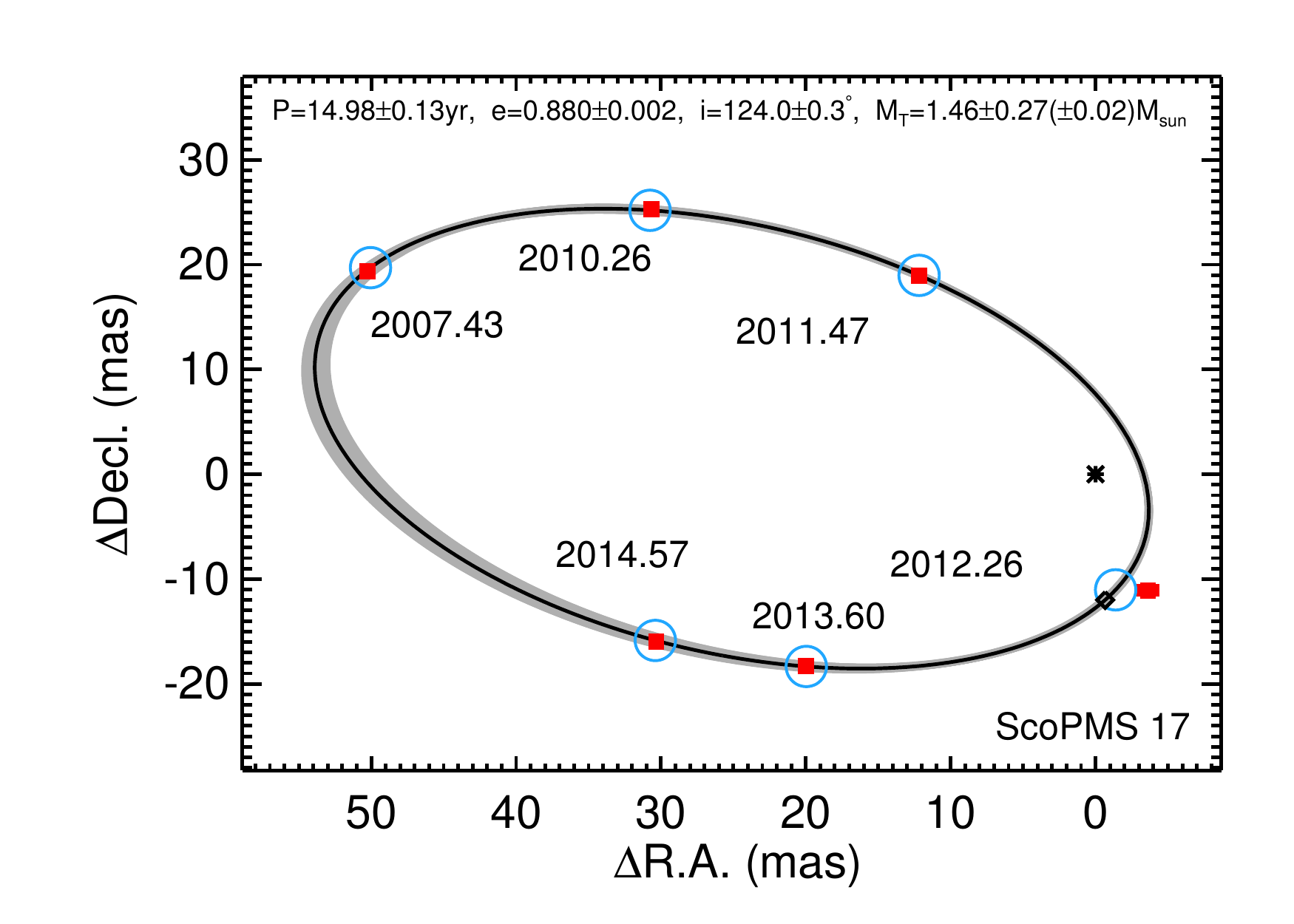}\\
\includegraphics[width=0.5\textwidth]{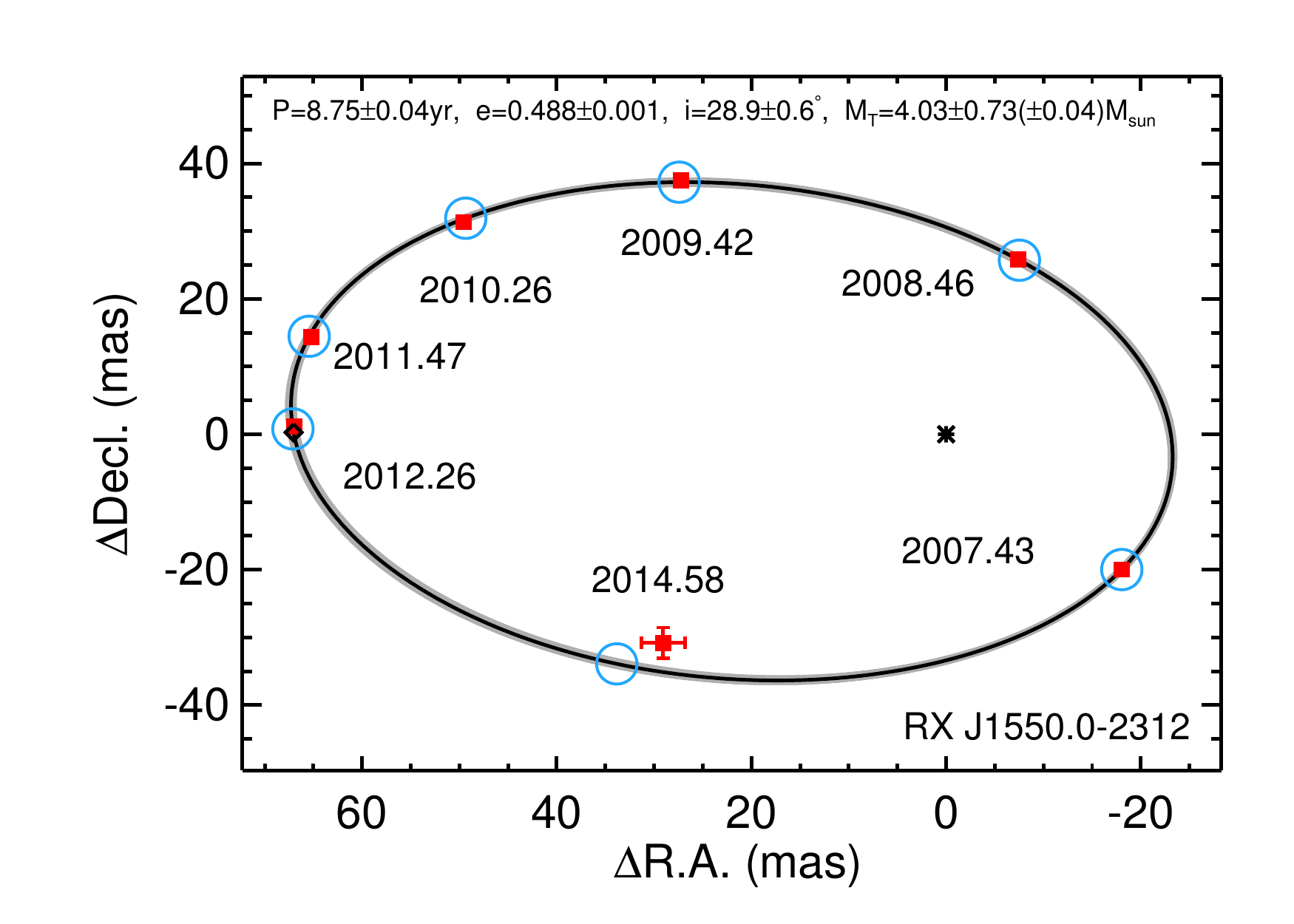}
\includegraphics[width=0.5\textwidth]{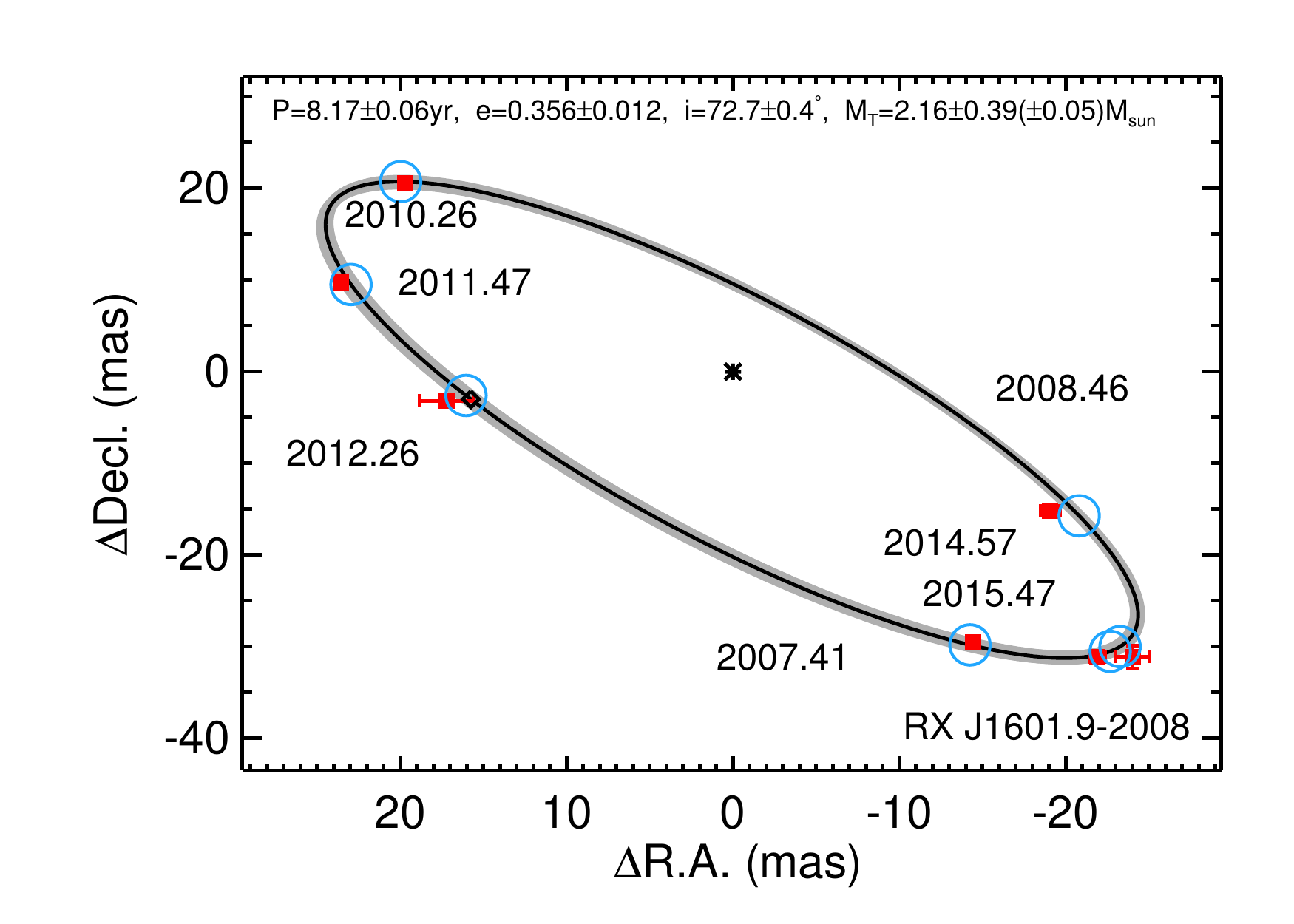}
\caption{As in Figure \ref{roxs47a_orbfit}, but for the orbits of six Upper Scorpius binaries, derived from NIRC2 AO aperture masking. The data is shown as red squares, with corresponding model fits shown as blue circles. We show the best fit orbit in black, with a 1-$\sigma$ region shaded in grey. The black diamond shows the orbital position of the secondary at the time of the corresponding HST observations (see Table \ref{hstphot_delta}). We also display the orbital period, semimajor axis, eccentricity and system mass at 145\,pc. The two uncertainties listed for the mass are derived by either including or excluding the $\pm$15\,pc distance uncertainty associated with Upper Scorpius membership. The bracketed uncertain, which only includes the uncertainty on the orbital parameters is representative of the system mass precision attainable with future GAIA parallaxes.}
\label{largeorbs}
\end{figure*}

The astrometric measurements taken in our orbit monitoring program were used to determine orbital solutions using a $\chi^2$ minimization on a grid of orbital parameters. We first generated random sample of 10$^4$ sets of semimajor-axis, eccentricity and system mass values. The range over which the semimajor-axis samples were taken spanned 0.5-1.5 times the maximum observed separation of the each binary system, while the eccentricity was sampled in the range $0 < e < 1$. We chose the system mass sample range based on the spectral types in Table \ref{targets}, with a conservatively large range of possible masses spanning 0.1 to 3\,\Msun. From each random system mass and semimajor-axis pair, we calculate the corresponding period, using the characteristic distance of Upper Scorpius according to Kepler's law $M_{\rm tot}=a^3P^{-2}(d_{US}^3)$. Using this random sample, we then fit the remaining four orbital parameters. We then further restrict the possible range of the randomly selected parameters and draw a new random sample centered on the current best fit values, and the process is repeated. The best fit parameters from this second sampling are then taken as the starting-point for a Monte-Carlo Markov-Chain process to more precisely determine the best fit orbital parameters.

Table \ref{orb_params} lists the orbital solutions for the stars in our program, while the corresponding orbital plots can be found in Figures \ref{roxs47a_orbfit} and \ref{largeorbs}. Typically, the semimajor axis (in angular units) and the period, the two important parameters for estimation of the system dynamical mass, are determined to better than $\sim$2--3\,\%. In the last two columns of Table \ref{orb_params}, we list the system dynamical masses at a fixed parallax of 7.5\,mas, chosen to represent the mean distance to the Upper Scorpius subgroup. These system masses can then be appropriately scaled to an alternate parallax $\pi_{n}$ through multiplication by $(7.5/\pi_{n})^3$, with appropriately adjusted uncertainties.

One of the tightest orbits that we have determined has also been identified as a single lined spectroscopic binary by \citet{guenther07}. The RV orbit has a period of 2045$\pm$16\,days and eccentricity of 0.2$\pm$0.03 which are very similar to our orbital period measurement of 1998.4$\pm$18.3\,days and eccentricity of 0.219$\pm$0.014. Because the distance to these stars is unknown, the mass ratio of the primary and secondary components cannot be directly disentangled from the combination orbital solutions, however the mass function from the RV orbit, $f_{m}=0.0049\pm0.0005$, can be included as additional information with which to fit models to the orbital and photometric data.

%%%%ORBITAL ELEMENTS TABLE%%%%%%%%%%%%%%%%%%%%%%
\begin{deluxetable*}{ccccccccc}
\tabletypesize{\footnotesize}
\tablewidth{0pt}
\tablecaption{Astrometric Orbital Elements}
\tablehead{
Star&\colhead{T$_0$}&\colhead{P}&\colhead{a}&\colhead{$\epsilon$}&\colhead{$\Omega$}&\colhead{$\omega$}&\colhead{$i$}&\colhead{M$_{\mathrm{tot}}$}\\
&\colhead{(JD)}&\colhead{(days)}&\colhead{(mas)}&&\colhead{($^\circ$)}&\colhead{($^\circ$)}&\colhead{($^\circ$)}&\colhead{(M$_\odot$)}
}
GSC 6209-735       & 2455182.8$\pm$3.5 & 2015.5$\pm$12.6 & 28.22$\pm$0.47 & 0.210$\pm$0.012 & 207.7$\pm$1.1 & 27.3$\pm$1.8   & 64.7$\pm$1.1   & 2.25$\pm$0.11($\pm$0.41)\\
GSC 6794-156       & 2453837.3$\pm$4.9 & 4851.4$\pm$29.8 & 58.19$\pm$0.31 & 0.262$\pm$0.004 & 328.6$\pm$1.4 & 45.2$\pm$1.5  & 161.4$\pm$1.2  & 3.40$\pm$0.02($\pm$0.62)\\
J160517.9-202420   & 2456553.5$\pm$0.4 & 2146.2$\pm$3.4  & 21.88$\pm$0.38 & 0.907$\pm$0.004 & 160.4$\pm$5.7 & 59.3$\pm$5.8   & 154.8$\pm$2.0  & 0.92$\pm$0.05($\pm$0.17)\\
ScoPMS 17          & 2455945.3$\pm$1.0 & 5471.5$\pm$47.1 & 47.58$\pm$0.37 & 0.880$\pm$0.002 & 199.6$\pm$0.3 & 286.1$\pm$0.3  & 124.0$\pm$0.3  & 1.46$\pm$0.02($\pm$0.26)\\
RX J1550.0-2312    & 2454412.5$\pm$0.5 & 3195.7$\pm$12.9 & 46.58$\pm$0.06 & 0.488$\pm$0.001 & 241.6$\pm$0.8 & 30.5$\pm$1.0   & 29.1$\pm$0.6   & 4.03$\pm$0.04($\pm$0.73)\\
RX J1601.9-2008    & 2454974.5$\pm$3.2 & 2983.3$\pm$20.5 & 36.18$\pm$0.27 & 0.356$\pm$0.012 & 223.6$\pm$0.5 & 106.9$\pm$0.7  & 72.7$\pm$0.4   & 2.16$\pm$0.05($\pm$0.38)\\
ROXs 47A           & 2455929.8$\pm$1.9 & 3007.1$\pm$42.9 & 32.69$\pm$0.35 & 0.818$\pm$0.009 & 24.1$\pm$11.0 & 243.0$\pm$10.9 & 16.3$\pm$2.0   & 1.57$\pm$0.03($\pm$0.29)\\
\enddata
\tablecomments{List of orbital elements and corresponding uncertainties fitted to the astrometric data for the objects monitored in our program. The final columns contain the total system masses for our targets, derived from the orbital parameters at a fixed distance of 145\,pc (6.9\,mas parallax), chosen to represent the mean distance of Upper Scorpius. The uncertainties provided are solely taken from the uncertainty in the orbital parameters, and the bracketed uncertainties include an uncertainty of $\pm$0.7\,mas on the parallax. Note that the large system mass for RXJ1550.0-2312 (M-type) implies that the star is very likely to be in the foreground with respect to Upper-Scorpius, with a parallax greater than $\sim$10\,mas.}
\label{orb_params}
\end{deluxetable*}

\begin{deluxetable*}{cccccccc}
\tabletypesize{\footnotesize}
\tablewidth{0pt}
\tablecaption{HST/WFC3 Unresolved Photometry}
\tablehead{
\colhead{Filter}   & \colhead{RXJ1550...}      & \colhead{RXJ1601...}      & \colhead{USco J1605...}    & \colhead{GSC6209-735}    & \colhead{GSC6794-156}     & \colhead{ROXs47A}         & \colhead{ScoPMS017}       \\
}
\startdata
F225W    &  ...             & 14.96$\pm$0.04   &  ...             &  ...             & 14.34$\pm$0.03   &  ...             &  ...             \\
F275W    & 18.43$\pm$0.14   & 13.48$\pm$0.03   & 18.19$\pm$0.12   & 15.43$\pm$0.04   & 12.88$\pm$0.02   & 18.99$\pm$0.18   & 18.08$\pm$0.12   \\
F336W    & 17.05$\pm$0.05   & 11.77$\pm$0.02   & 17.16$\pm$0.06   & 13.20$\pm$0.02   & 11.15$\pm$0.02   & 17.29$\pm$0.06   & 16.66$\pm$0.05   \\
F390W    & 16.47$\pm$0.03   & 11.68$\pm$0.02   & 16.60$\pm$0.03   & 13.01$\pm$0.02   & 11.07$\pm$0.02   & 16.48$\pm$0.03   & 16.17$\pm$0.03   \\
F395N    & 16.67$\pm$0.06   & 12.15$\pm$0.02   & 16.77$\pm$0.06   & 13.50$\pm$0.03   & 11.54$\pm$0.02   & 16.90$\pm$0.07   & 16.41$\pm$0.06   \\
F438W    & 15.88$\pm$0.03   & 11.41$\pm$0.02   & 15.98$\pm$0.03   & 12.58$\pm$0.02   & 10.77$\pm$0.02   & 15.72$\pm$0.03   & 15.56$\pm$0.03   \\
F467M    &  ...             & 11.07$\pm$0.02   &  ...             &  ...             & 10.39$\pm$0.02   &  ...             &  ...             \\
F475W    & 15.03$\pm$0.02   &  ...             & 15.14$\pm$0.02   & 12.05$\pm$0.02   &  ...             & 14.80$\pm$0.02   & 14.81$\pm$0.02   \\
F547M    &  ...             & 10.43$\pm$0.02   &  ...             &  ...             & 9.76$\pm$0.02    &  ...             &  ...             \\
F555W    & 14.28$\pm$0.02   &  ...             & 14.35$\pm$0.02   & 11.62$\pm$0.02   &  ...             & 13.91$\pm$0.02   & 14.05$\pm$0.02   \\
F625W    & 13.25$\pm$0.02   &  ...             & 13.34$\pm$0.02   & 10.87$\pm$0.02   &  ...             & 12.83$\pm$0.02   & 13.06$\pm$0.02   \\
F631N    &  ...             & 9.96$\pm$0.02    &  ...             &  ...             & 9.30$\pm$0.02    &  ...             &  ...             \\
F656N    & 12.04$\pm$0.03   & 9.39$\pm$0.02    & 12.13$\pm$0.03   & 10.30$\pm$0.02   & 8.67$\pm$0.02    & 11.71$\pm$0.03   & 11.86$\pm$0.03   \\
F673N    &  ...             & 9.73$\pm$0.02    &  ...             &  ...             & 9.09$\pm$0.02    &  ...             &  ...             \\
F775W    & 11.69$\pm$0.02   & 9.38$\pm$0.02    & 11.95$\pm$0.02   & 10.25$\pm$0.02   & 8.82$\pm$0.02    & 11.47$\pm$0.02   & 11.73$\pm$0.02   \\
F850LP   & 10.85$\pm$0.02   & 8.98$\pm$0.02    & 11.14$\pm$0.02   & 9.85$\pm$0.02    & 8.36$\pm$0.02    & 10.56$\pm$0.02   & 10.92$\pm$0.02   \\
\enddata
\tablecomments{Hubble Space Telescope, Wide Field Camera 3  unresolved photometry for the binary systems in our orbit monitoring program. The star names have been abbreviated, but can be found in full in Table \ref{targets}. We have included the aperture photometry for the narrow-band filters, which were not used in the following analysis, for completeness and future use.}
\label{hstphot}
\end{deluxetable*}

\section{Hubble Space Telescope Observations}
In addition to AO imaging, we have obtained single-epoch observations of these binary systems with the Hubble Space Telescope (HST) Wide Field Camera 3 (WFC3), in a variety of visible filters spanning wavelengths of $200-1000$\,nm. We took either three or four exposures in each filter, to which was applied the standard HST reduction, calibration and cosmic ray rejection procedure \citep{wfc3dhb}. We calculated binary system unresolved magnitudes for each WFC3 filter using aperture photometry on the HST drizzle images with a 0.4" radius star aperture and a sky annulus of $4-6$". We used a 0.4" aperture to facilitate calibration with the readily available WFC3 filter information, and to ensure that the wide companion in the ROXs 47A triple system was excluded from the both the star aperture and the sky subtraction region. For GSC 6794-156, the images had the centre of the star PSF flagged (most likely incorrectly) as cosmic rays in all three uncombined FLT images, and so for this star the combined drizzle images were expected to give incorrect photometry. For this binary system we applied the aperture photometry procedure to the uncombined FLT images.

Given the orbital solutions we have determined for these systems, we can accurately predict the separation and position angle of each binary system at the epoch of observation with the HST, to within a few milliarcseconds. Combined with the stability of the HST point spread function (PSF), we can derive differential photometry from our observations even though the binary separations are typically $<60$\,mas. Our PSF fitting is modeled after our previous work \citep{garcia2015}, with the addition of a further complication due to the short exposure times used in many of the filters in which our sample was observed. For the majority of the observations, we used three exposure which were often shorter than 1\,s, an exposure regime in which vibrations induced by the WFC3 shutter causes observable blur in the images. It has been shown in on-sky and laboratory test that following the rotation of the WFC3 shutter at the beginning of an exposure, there is a period of a few seconds in which the PSF shape is degraded due to vibrations in the camera \citep{hartig_shutter}. In order to account for the presence of the vibrational blurring on the output contrast ratios, we both employed a large library of unblurred PSF standards, and characterized the effect of the vibration on the photometry of a binary system. Firstly, we used a large library of archival data with long exposures to identify multiple unblurred PSF reference sources in each filter of interest for both the C512C and M512C subs arrays of the UVIS2 detector. Using the Tiny Tim software \citep{tinytimproceedings}, we created PSF models for the WFC3 filters, these were then fit to the PSF standards to determine a modified PSF that most closely fits the large sample of reference data. This process involved vetting the PSF references for obvious $>$1\,pixel binaries and sources where cosmic rays landed within a few pixels of the target center. All other contaminants outside of a four pixel radius from the PSF centre were handled automatically via sigma clipping with respect to the residual distribution of the entire PSF reference library.  

\begin{figure}
\centering
\includegraphics[width=0.45\textwidth]{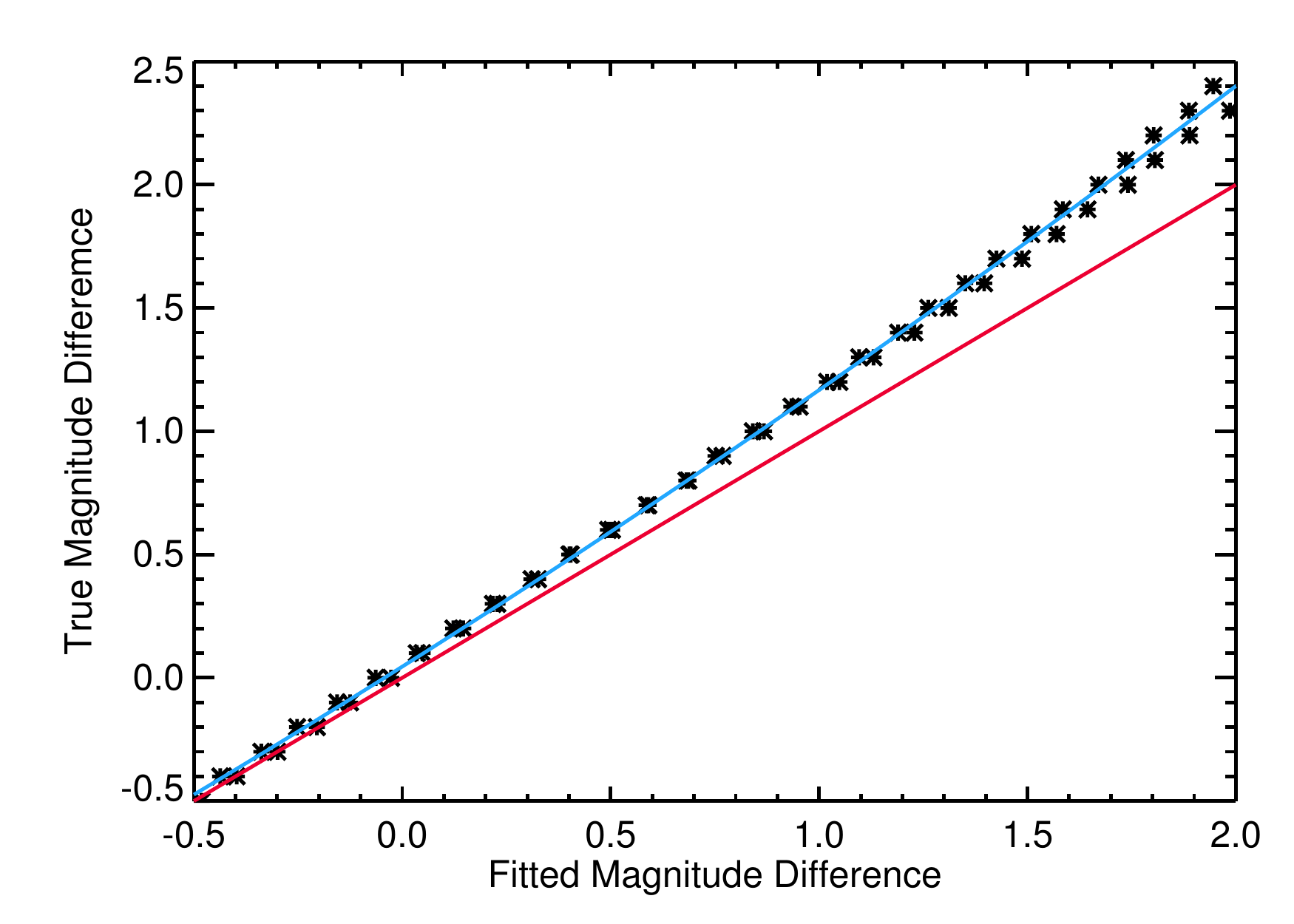}
\caption{Contrast ratio correction for a synthetic binary separation $(x,y)=-(2,2)$\,pixels on the WFC3 detector. The black stars are the fitted contrast ratios, the blue line the polynomial fit used to correct the actual binary system data, and the red line is the case for a zero correction.}
\label{contrast_correction}
\end{figure}

We created a set of synthetic, blurred binary systems at different integer-pixel separations and contrast ratios, using multiple images of ROXs 47B, the wide tertiary companion to ROXs 47A which was in the frame of the ROXs 47A images. We then convolved the reference model PSF in each filter, created as described above, with a Gaussian kernel and attempted to recover the original input contrast ratios. To do this we fixed the separation and P.A. of the synthetic binaries, and allowed the position of the primary star, flux normalization, flux ratio and blur FWHM to vary. With this method, we were able to create a grid of photometry corrections at different separations for both the mild and severely blurred images, by fitting polynomials as a function of output contrast ratios. These corrections were then applied to the PSF fits for the binary systems in our sample by interpolating on a 3D-grid of position and blur. Figure \ref{contrast_correction} displays an example contrast ratio correction for one of the synthetic binary systems. The uncertainties for the magnitude differences was taken as the RMS between the images in each filter and the uncertainty on the correction combined in quadrature. The aperture and differential photometry for our sample is presented in Tables \ref{hstphot} and \ref{hstphot_delta} respectively.

\begin{figure}
\centering
\includegraphics[width=0.45\textwidth]{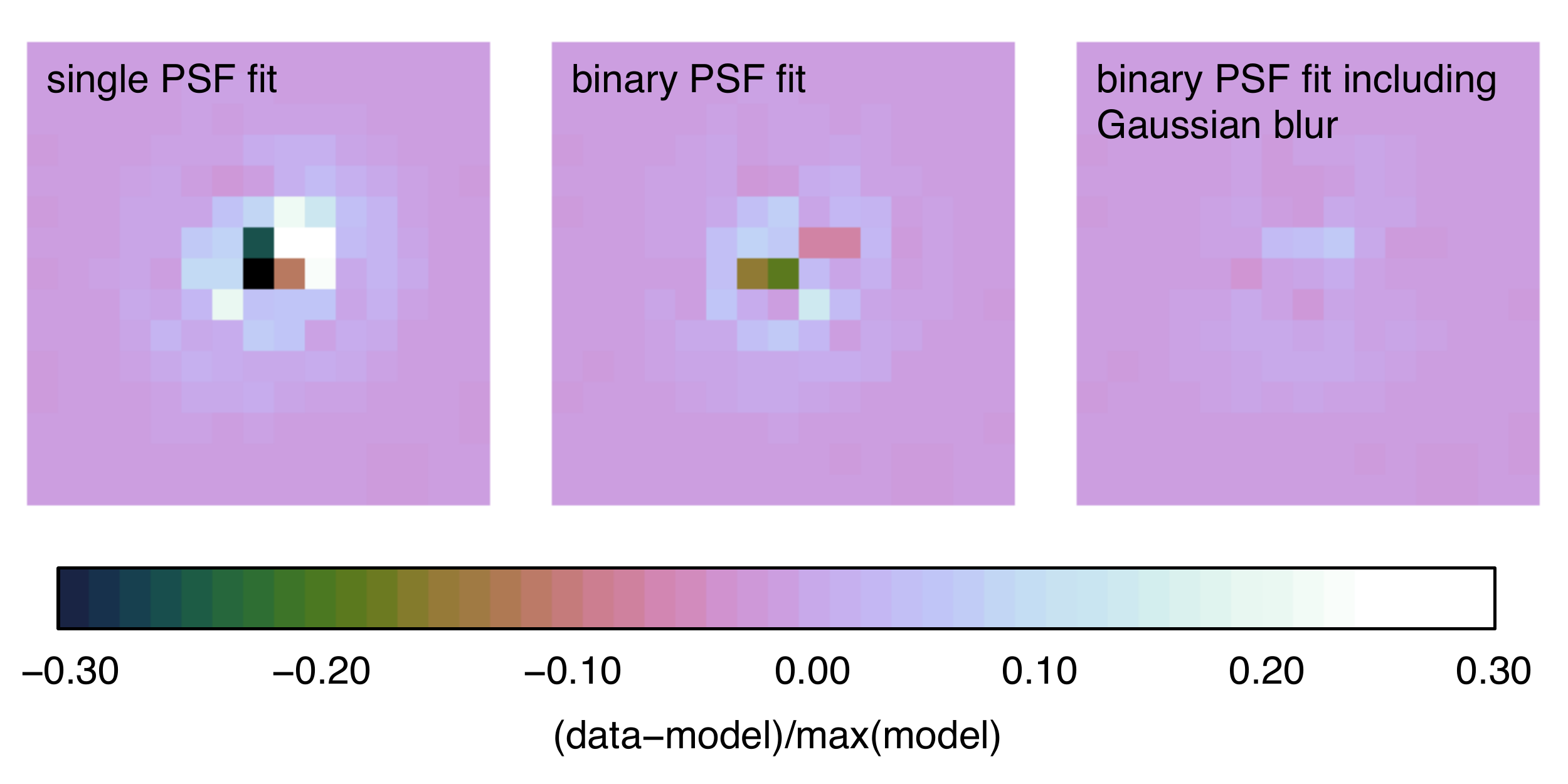}
\caption{Example PSF fit residuals for the binary system RXJ-1550.0-3212, in the filter F775W and the M512C subarray. From left to right panels show the residuals when fitting a single PSF, binary PSF, and binary PSF with gaussian blur to account for the shutter vibration. At the epoch of this HST observations, thie binary components were separated by 65.8\,mas at a position angle of 90.1$^\circ$ ($dx,dy=1.46,0.8$ pix). }
\label{psf_resid}
\end{figure}
%%RXJ 1550.0-3212
%F775W
%M512C
%imfile = ibsl23j8q
%blur = 1.242369
%PA = 90.1, sep = 65.8
%%dx = 1.46 dy = 0.804

\begin{deluxetable}{cccc}
\tabletypesize{\footnotesize}
\tablewidth{0pt}
\tablecaption{HST/WFC3 Resolved Photometry}
\tablehead{
\colhead{Property}  & \colhead{RXJ1550...}       & \colhead{USco J1605...}   & \colhead{GSC6794-156}\\
}
\startdata
MJD                                     &     56031           &   56032        & 56032\\
$\rho$\,(mas)                     &  65.8$\pm$0.2   &  31.2$\pm$0.7  & 71.8$\pm$0.4     \\  
P.A.\,($^\circ$)                    &   90.1$\pm$1.2   & 270.6$\pm$8.5 & 115.1$\pm$1.8  \\ 
$\Delta$F225W    &  ...                         &  ...                        & 1.71$\pm$0.35                            \\
$\Delta$F275W    & ...                          &  ...                        & 1.29$\pm$0.31                    \\
$\Delta$F336W    & 1.41$\pm$0.33  & ...                        & 0.94$\pm$0.31               \\
$\Delta$F390W    & 1.68$\pm$0.25  &  ...                       & 0.75$\pm$0.27          \\
$\Delta$F395N    & 1.49$\pm$0.36   &  ...                          & 0.81$\pm$0.48           \\
$\Delta$F438W    & 1.76$\pm$0.38   &  ...                        & 0.70$\pm$0.49    \\
$\Delta$F467M    &  ...                          &  ...                        & 0.46$\pm$0.23                 \\
$\Delta$F475W    & 2.31$\pm$0.17  &  ...                         &  ...                                           \\
$\Delta$F547M    &  ...                          &  ...                        & 0.53$\pm$0.17                    \\
$\Delta$F555W    & 1.85$\pm$0.22   & 0.52$\pm$0.40 &  ...                                   \\
$\Delta$F625W    & 1.60$\pm$0.15  &  0.66$\pm$0.21 &  ...                          \\
$\Delta$F631N    &  ...                          &  ...                        & 0.44$\pm$0.06              \\
$\Delta$F656N    & 1.11$\pm$0.04   & 0.71$\pm$0.14 & 0.55$\pm$0.07       \\
$\Delta$F673N    &  ...                          &  ...                         & 0.45$\pm$0.06              \\
$\Delta$F775W    & 1.10$\pm$0.09  & 0.68$\pm$0.18  & ...      \\
$\Delta$F850LP   & 0.95$\pm$0.02  & 0.84$\pm$0.33  & 0.52$\pm$0.06         \\
\enddata
\tablecomments{Hubble Space Telescope, Wide Field Camera 3 differential photometry for the binary systems in our orbit monitoring program. The star names have been abbreviated, but can be found in full in Table \ref{targets}. The first two rows list the separation and position angle of the companions at the epoch of HST observation. In general we have not reported any magnitude difference with an uncertainty larger than 0.5 magnitudes, or where the PSF fit was determined to have failed due to the close proximity of the two components. Magnitude differences were not derivable for the binary systems ScoPMS 17, GSC6209-735, RX J1601.9-2008 and ROXs 47A, due to the high contrast or sub-pixel proximity of the components of these systems at the time of HST observations (12, 33.4, 16, 39\,mas respectively).}
\label{hstphot_delta}
\end{deluxetable}

%%%%HST PHOT TABLE %%%%%%%%%%%%%%

\section{Binary SED Properties}
\begin{deluxetable*}{ccccccccc}
\tabletypesize{\footnotesize}
\tablewidth{0pt}
\tablecaption{Model System Properties}
\tablehead{
\colhead{Name} & \colhead{T$_{eff,p}$} & \colhead{T$_{eff,s}$} & \colhead{L$_p$} & \colhead{L$_s$} & \colhead{Age} & \colhead{M$_{p}$} & \colhead{M$_{s}$} & \colhead{D} \\    
    & \colhead{(K)}  & \colhead{(K)} & \colhead{(L$_\odot$)} & \colhead{(L$_\odot$)} & \colhead{(Myr)} & \colhead{(M$_\odot$)} &  \colhead{(M$_\odot$)} & \colhead{(pc)}
}
\startdata
GSC 6794-156    & 5700$\pm$201    & 5340$\pm$114 & 3.544$\pm$0.594 & 1.977$\pm$0.376 & 9$\pm$3 & 1.54$\pm$0.13 & 1.42$\pm$0.11 & 138$\pm$15\\
GSC 6209-735    & 5140$\pm$178    & ...                        & 1.283$\pm$0.258 & ...                       & 12$\pm$6 & 1.23$\pm$0.14 &  ...                      & 133$\pm$16\\
RX J1601.9-2008 & 5580$\pm$152    &  ...                      & 2.777$\pm$0.506 & ...                       & 9$\pm$3 & 1.51$\pm$0.13 &  ...                       & 141$\pm$16\\
USco J1605...   & 3760$\pm$50     & 3460$\pm$69 & 0.213$\pm$0.033 & 0.138$\pm$0.024 & 6$\pm$1 & 0.71$\pm$0.05 & 0.57$\pm$0.05 & 161$\pm$16\\
ScoPMS 17       & 3860$\pm$36     & 3300$\pm$58 & 0.301$\pm$0.045 & 0.132$\pm$0.022 & 7$\pm$2 & 0.72$\pm$0.05 & 0.48$\pm$0.05 & 136$\pm$15\\
RX J1550.0-2312 & 3690$\pm$38     & 3010$\pm$67 & 0.316$\pm$0.048 & 0.138$\pm$0.024 & 16$\pm$3 & 0.71$\pm$0.05 & 0.35$\pm$0.05 & 93$\pm$15\\
ROXs 47A        & 3850$\pm$36     & 3650$\pm$39 & 0.435$\pm$0.066 & 0.343$\pm$0.052 & 4$\pm$1 & 0.71$\pm$0.05 & 0.62$\pm$0.05 & 137$\pm$15\\
\enddata
\tablecomments{System properties derived from fitting BT-Settl models to the binary system photometry and comparison. Due to the high contrast ratios for RX J1601.9-2008 and GSC6209-735, only the primary temperature and luminosity was determined. The ages and masses are taken from comparison to the Padova isochrones.}
\label{bsedpars}
\end{deluxetable*}

\begin{figure}
\includegraphics[width=0.5\textwidth]{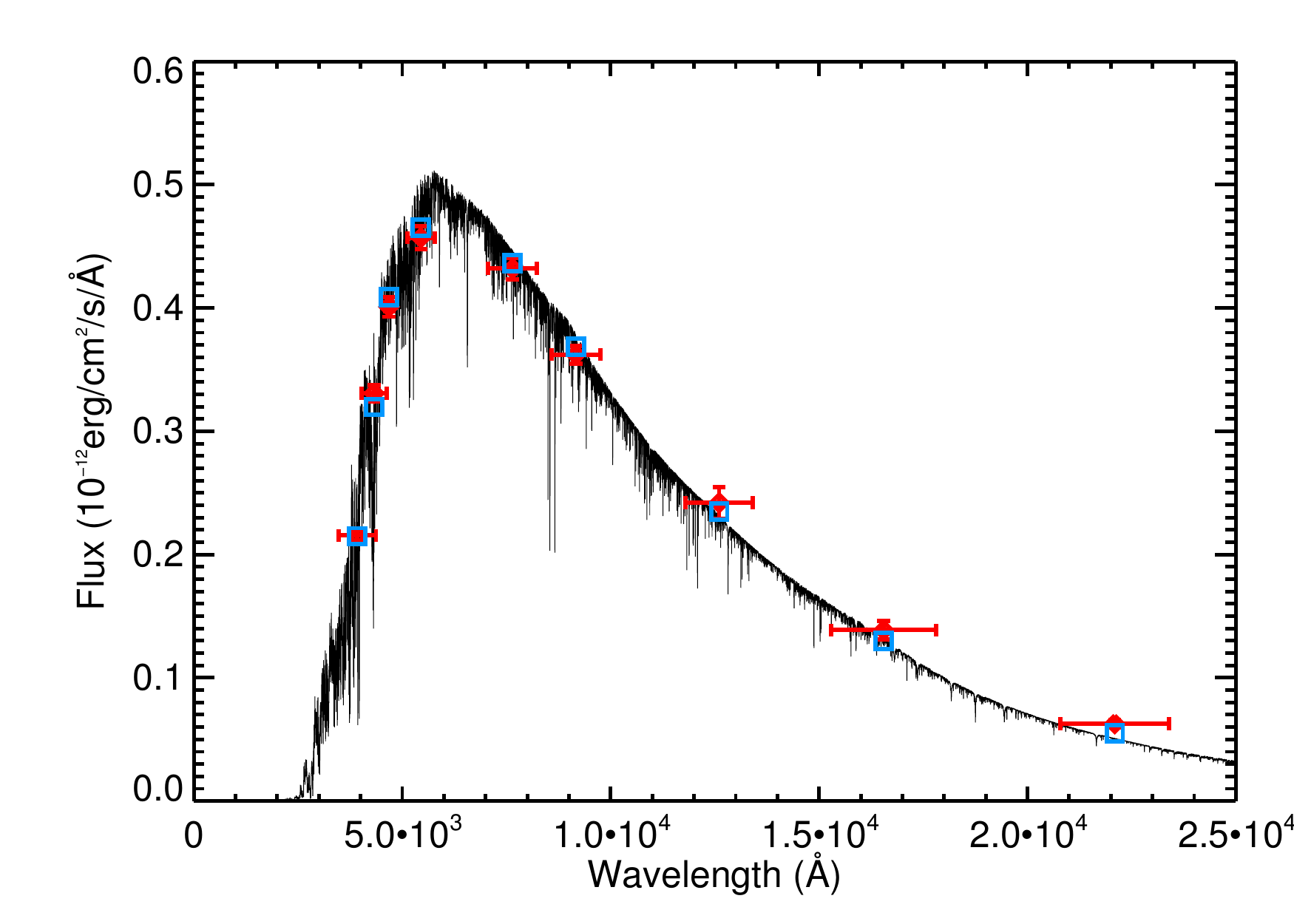}\\
\includegraphics[width=0.5\textwidth]{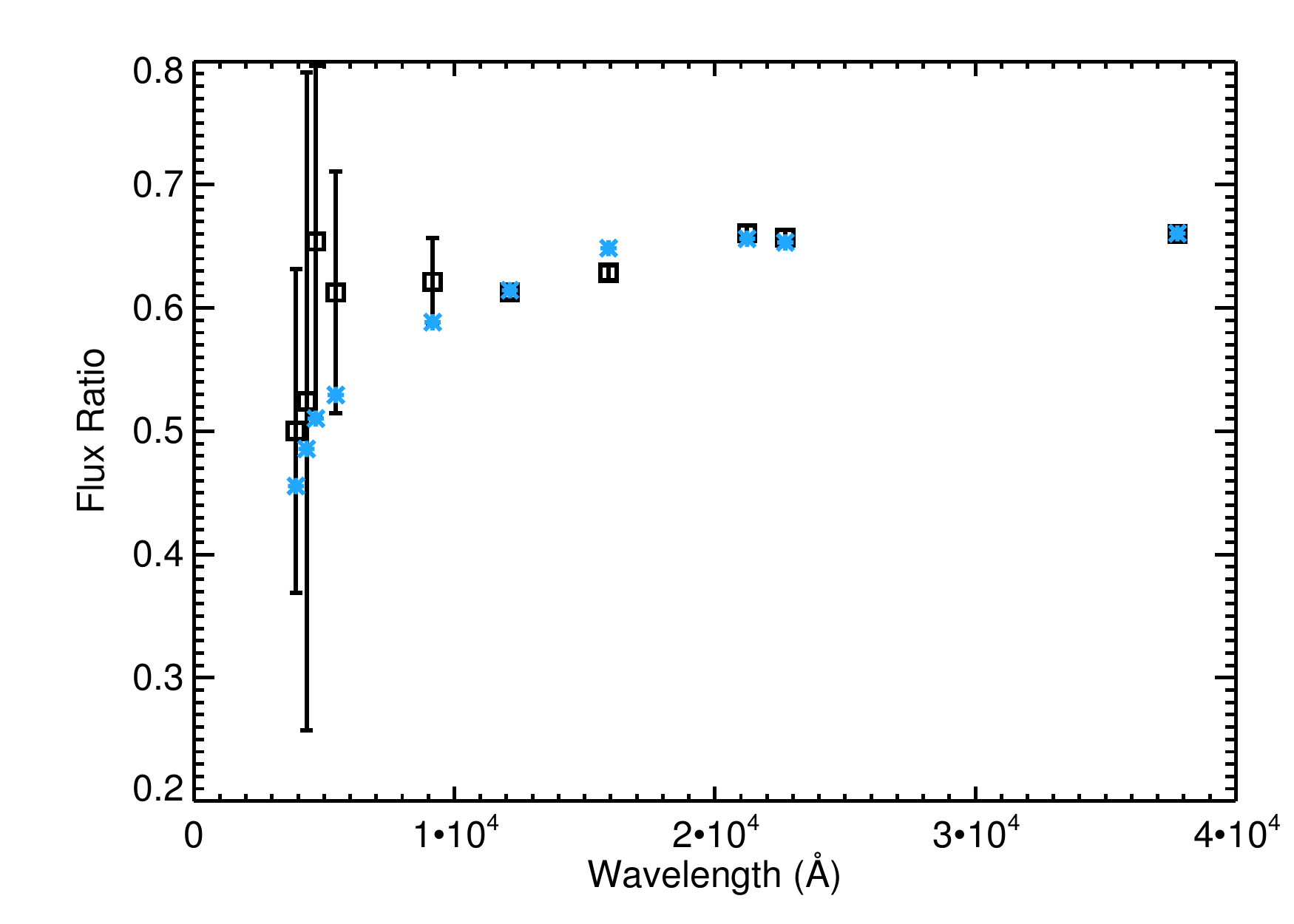}
\caption{Binary SED fitted unresolved photometry (a) and flux ratios (b) for GSC 6794-156. The Black line is the combined BT-Settl atmospheres at the two best-fit temperatures, the red points with error bars are the observed photometry, and the blue squares are the model photometry. The best fit atmosphere shown here is a combination of two atmospheres with temperatures 5700\,K and 5350\,K.}
\label{binary_sed_p02065}
\end{figure}

We have used the unresolved WFC3 and 2MASS photometry, combined with the resolved magnitude differences from the NIRC2 aperture masking observations to determine temperatures and luminosities for the components of the binary systems in our monitoring sample. We employed a grid of paired BT-Settl \citep{allard95,baraffe15} synthetic spectra, convolved with filter profile for each of the WFC3 and 2MASS filters, and compared these to the observed values using a $\chi^2$ methodology. The is some degeneracy between extinction and temperature, so we allowed A$_V$ to vary from 0.6-1.2 (0.24$<$E(B-V)$<$0.36), which is a typical range of values for Upper Scorpius members \citep{preibisch02,wifes1_2015}. For ROXs 47A, the single Ophiuchus member in our sample, we doubled the prescribed reddening in anticipation of stronger extinction towards the star-forming region. To determine the component temperatures, we take both BT-Settl atmospheres and combine them at a given ratio, and then scale the resulting combined spectra to best match the observed photometry, repeating the process for a generous range of effective temperatures. Figure \ref{binary_sed_p02065} displays the SED fit for GSC 6794-156 as an example. The primary and secondary bolometric fluxes are then calculated for the individual best fit temperatures, excluding the effects of extinction. With the exception of GSC 6209-735 and RX J1601.9-2008, the two highest contrast binary systems in our sample, we were able to determine individual component temperatures using this method. The uncertainty in the derived temperatures and bolometric fluxes are both strongly dominated by the unconstrained extinction.

\begin{figure}
\includegraphics[width=0.5\textwidth]{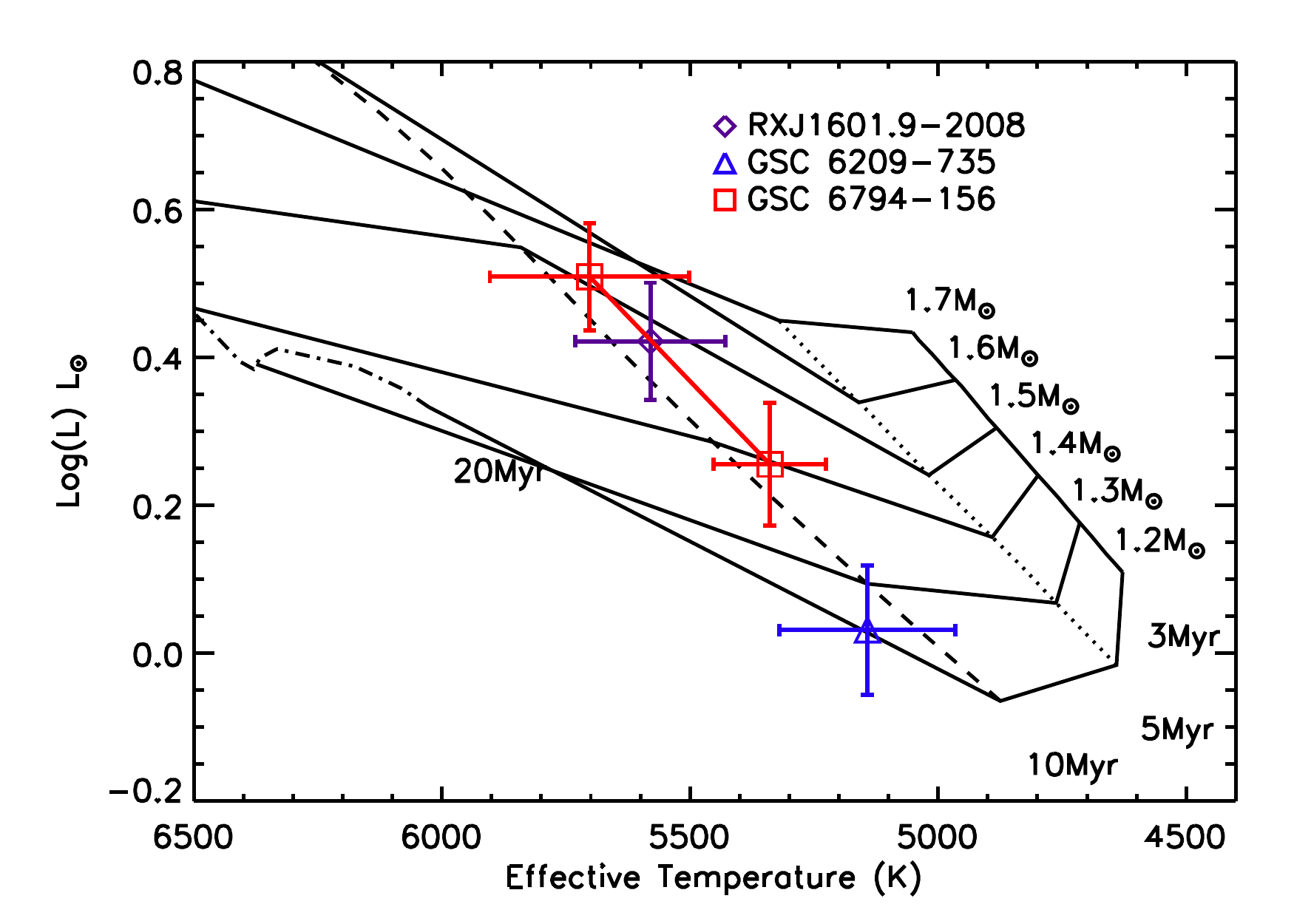}\\
\includegraphics[width=0.5\textwidth]{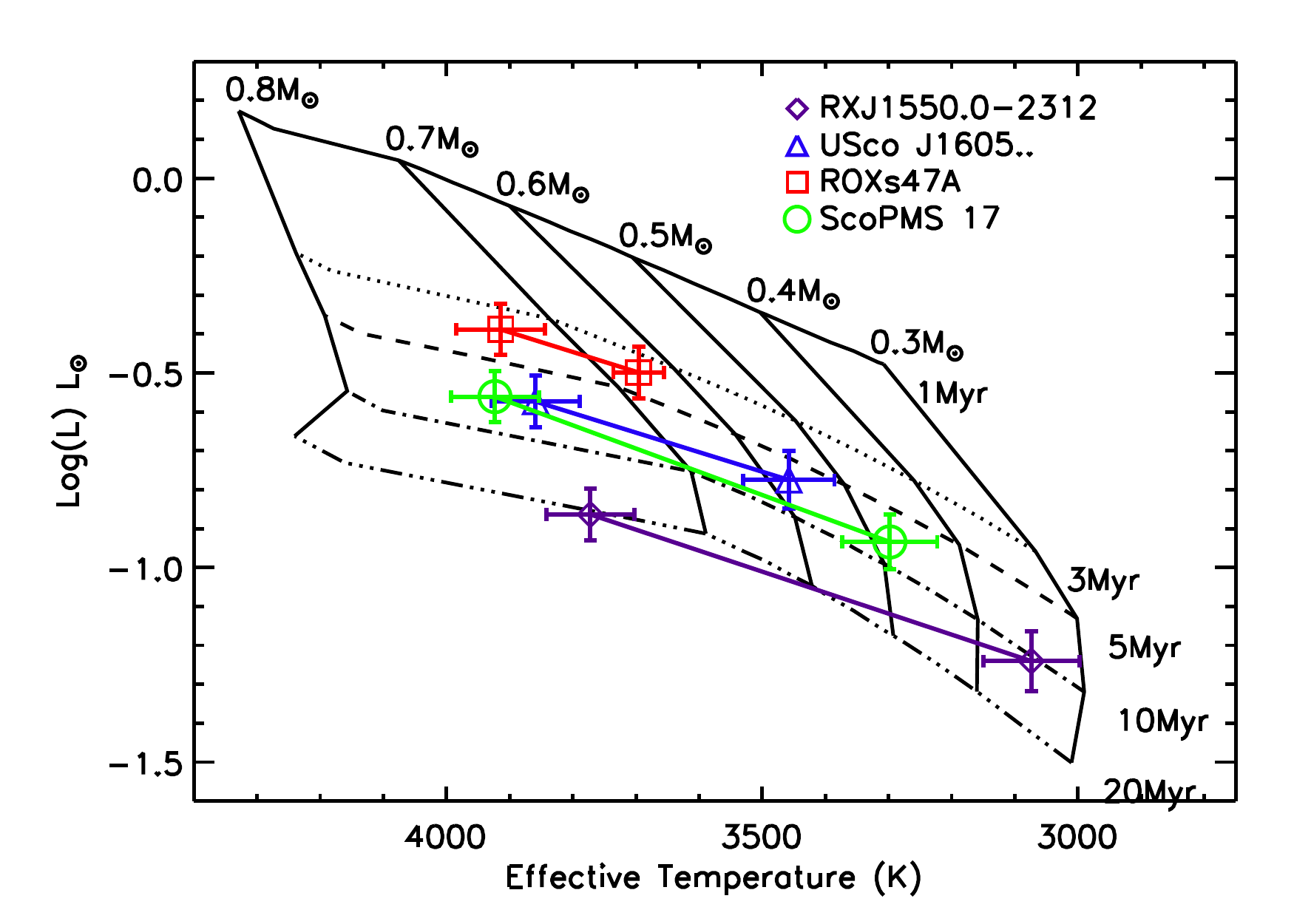}
\caption{HR-diagram positions for the binary system components in our orbit monitoring sample. Components of each binary system are displayed in a different color and with a different symbol, and is connected to the corresponding companions via a colored line. The black lines indicate the isochrone and isomass lines taken from the latest Padova models \citep{chen14_padova}. For clarity, we display the higher mass binaries in panel (a) and the M-type binary systems in panel (b).}
\label{gmhrd}
\end{figure}

Given these estimates for the component fluxes and temperatures, we can now place these PMS stars on the HR-diagram and derive age and mass estimates from the data. Upper Scorpius inhabits a significant volume of space, with a depth of $\pm$15\,pc \citep{zeeuw99,myfirstpaper}. This comprises a significant uncertainty in determining HR-diagram luminosities for the binary systems in our sample. However, the system dynamical mass observable , derived from the binary system orbital parameters, provides an indirect constraint on the distance of the binary system. As mentioned above, the dynamical mass observable for an astrometric orbit $M_{\rm tot}d^{-3}=a^3P^{-2}$ is strongly dependent on the binary system distance, and can be used to better constrain the HR-diagram position of the binary system components. For each system we then fit for distance and component masses as a function of the estimated temperatures and luminosities, and the measured dynamical mass observable. For the two systems with unconstrained secondary luminosity and temperature, we conservatively fix the secondary mass in the fit to be 0.5$\pm$0.3\,M$_\odot$ and fit for the primary mass and system distance.  Figure \ref{gmhrd} displays the HR-diagrams for both the G/K and M-type binary systems in our sample, and the resulting properties of the components are listed in Table \ref{bsedpars}. 

The two G-type members, GSC 6794-156, GSC 6209-735, and RXJ 1601.9-2008, all have HR-diagram ages of $\sim$10\,Myr, which agrees directly with the most recent age measurement of \citet{pecaut12}. The M-type binaries ScoPMS 17 and USco J160517.9-202420 both appear slightly younger than the G-type binaries, with HR-diagram ages of $\sim$7\,Myr. We note that RXJ 1550.0-2312 appears significantly older than the other M-type binary system in our sample, with and age more consistent with Upper Centaurus Lupus membership.

\section{Bayesian Age Estimation}

While the SED fitting and HR-diagram position estimates for the binary component properties produce useful results in characterizing both the individual systems and the overall population in Upper Scorpius, the uncertainties introduced due to the unconstrained extinction warrant a more sophisticated approach that uses all available information to simultaneously determine the relevant parameters. The goal is to produce a method for determining the age, masses, distance and extinction for the stars in a binary system of known orbit, drawing data from the orbital parameters and measured magnitudes in a number of filters, as well as a contrast ratio in one or more filters. We phrase the problem in terms of Bayes' Theorem:

\begin{equation}
P(\Phi | D) \propto P(\Phi)P(D | \Phi),
\label{btheorem}
\end{equation}

where $\Phi$ represents a model and $D$ represents the data.  The model $\Phi$ consists of an age, model parallax ($\pi_m$), primary and secondary masses ($M_p$ and $M_s$), a reddening parameter ($E(B-V)$), and a set of isochrones, which map mass, reddening and age to magnitudes in different filters which can be compared to the data.

The data, $D$, consist of an association parallax in the absence of a directly measured distance ($\pi_{\star}$=$7.5\pm1.7$\,mas, taken from \citet{myfirstpaper}), the total mass observable ($M_T\pi^3 = a^3/P^2$) calculated from the orbital period and semimajor axis, the magnitude difference in one or more filters ($\Delta m_i$), taken from our AO aperture masking observations and HST photometry, and a set of unresolved magnitudes in available catalog filters $\{m_{\star, i}\}$, including \emph{APASS} BVgri filters, 2MASS J,H and K, and HST WFC3 filters. The treatment of the total mass observable  ($M_T\pi^3 = a^3/P^2$) and its corresponding uncertainty requires some care, because the fitter orbital period ($P$) and semimajor axis ($a$) are often highly correlated. To properly compute the uncertainty in the total mass observable, we take the full covariance matrix of the orbital parameters and use the coordinate transform to recast it in the total mass observable coordinates.

In the fortuitous case where both a visual orbit and a radial velocity orbit (either single or double lined) are available simultaneously for a single binary system, we can incorporate the additional radial velocity information, such as mass ratio for the double lined spectroscopic orbit, or the primary mass function ($f(M)$). Incorporation of the mass ratio into the following framework is trivial, however the inclusion of the mass function, as is the case for GSC 6209-735,  requires some description. We combine the mass function ($f(M)$) with the inclination taken from the visual orbit in order to define a mass function observable $M_{f(M)}$;
\begin{equation}
M_{f(M)} = \frac{M_s^3}{(M_p+M_s)^2} = \frac{f(M)}{\sin{i}^3}.
\end{equation}

To compute the Bayesian probability, we first expressed the models in terms of the more directly comparable parameters such as magnitude difference and unresolved magnitude using marginalization:

\begin{equation}
P(D | \Phi ) = P(D | \phi) P(\phi | \Phi),
\label{margin}
\end{equation}
where $\Phi=\{\mathrm{Age}, \pi_m, M_p, M_s,m_{p,i}, \mathrm{m}_{s,i}\}$,  $\phi=\{ \pi_m, m_{s+p,i}, \Delta m_{i}, (M_p+M_s)\pi_m^3,M_s^3(M_p+M_s)^{-2})\}$, and $\Delta m_i$ is the magnitude difference between secondary and primary in filter $i$ for the given primary and secondary masses.  $m_{s+p,i}$ is the unresolved magnitude in filter $i$ of the primary and secondary for the given masses and at the given parallax, reddened according to the \citet{savage_mathis79} extinction law and the model reddening parameter. We have included in this description the mass function observable because it is used for the binary system GSC 6209-735 for which we have a mass function measurement, however an additional term for the mass ratio can be easily added to the formality above.

Note that the above transformation is simple because the new parameters are directly given by the original model and so $P(\phi | \Phi) = 1$, leaving the following:

\begin{equation}
P(\phi | D) = \frac{P(D | \phi) P(\Phi)}{P(D)}.
\label{after_margin}
\end{equation}

For the prior probability distribution $P(\Phi)$, we have chosen a uniform distribution, meaning that all values of the model parameters are initially treated as being equally likely. For each set of model parameters, we then calculate $P(D | \phi)$, which takes the following form when separated into individual variables:
\begin{equation}
\begin{array}{c}
P(D|\phi) =  P(\pi_{\star}|\pi_m) P(M_t\pi^3|(M_{p+s})\pi_m^3)\\ 
~\\
P(M_{f(M)}|M_s^3M_{p+s}^{-2})\\
~\\
	\prod_i P(m_{\star,i} | m_{s+p, i}) \prod_j P(\Delta m_{\star,j} | \Delta m_{j}),
\end{array}
\label{breakdown}
\end{equation}

where $i$ and $j$ indicate multiplication over all available photometric filters. The probabilities in the above equations are modelled as normal distributions, with standard deviation given by the uncertainties in the data:

\begin{equation}
\begin{array}{c}
P(\pi_{\star} | \pi_m) \propto \exp{\left(-\frac{(\pi_{\star}-\pi_m)^2}{2\sigma_{\pi_{\star}}}\right)},\\
\\
P(M_t\pi^3|(M_{p+s})\pi_m^3) \propto \exp{\left(-\frac{(M_t\pi^3 - (M_p+M_s)\pi_m^3)^2}{2\sigma_{M_t\pi^3}^2}\right)},\\
\\
P(M_{f(M)}|M_s^3M_{p+s}^{-2}) \propto \exp{\left(-\frac{(M_{f(M)} - M_s^3(M_p+M_s)^{-2})^2}{2\sigma_{M_{f(M)}}^2}\right)},\\
\\
P(m_{\star,i} | m_{s+p, i}) \propto \exp{\left(-\frac{(m_{\star,i} - m_{s+p,i})^2}{2\sigma^2_{m_{\star,i}}}\right)},\\
\\
P(\Delta m_{\star,j} | \Delta m_{j}) \propto \exp{\left(-\frac{(\Delta m_{\star,j} - \Delta m_j)^2}{2\sigma^2_{\Delta m_{\star,j}}}\right)},\\
\end{array}
\label{main_eq}
\end{equation}

where we have omitted the usual normalization factors for the purpose of readability.
We then calculate the probability of a grid of model parameter sets using equations \eqref{breakdown} and \eqref{main_eq} to determine the most likely value of age, component masses and parallax. This was done with age ranging from $1-25$\,Myr in steps of 1\,Myr, and primary and secondary mass ranging from a minimum of 0.5\,\Msun to 2\,\Msun in steps of 0.01\,\Msun. The model parallax was varied from $2-15$\,mas in steps of $0.1$\,mas, and the reddening parameter $E(B-V)$ was varied from $0-1$ in steps of 0.05 magnitudes. Once a likely solution was found, we decreased the step size and sampling range to fully sample the probability distribution, using steps of 0.25 to 0.5\,Myr in age and 0.005\,\Msun in mass, and interpolating between isochrones. We also added in quadrature an error of 0.05 magnitudes to all non-simultaneous photometric measurements to account for the average variability of PMS dwarfs \citep{rotation_review2007}.

\begin{deluxetable*}{lccccc}
\tabletypesize{\footnotesize}
\tablewidth{0pt}
\tablecaption{APASS Photometry}
\tablehead{
\colhead{Star} & \colhead{B}&\colhead{V}&\colhead{g}&\colhead{r}&\colhead{i}
}
\startdata
RXJ1550.0-2312  		     & 15.613$\pm$0.479 & 14.065$\pm$0.054 & 14.760$\pm$0.337 & 13.342$\pm$0.196 & 12.059$\pm$0.102  \\
RXJ1601.9-2008		     & 11.333$\pm$0.043 & 10.380$\pm$0.030 & 10.985$\pm$0.187 & 10.086$\pm$0.004 & 9.637$\pm$0.011\\
USco J160517.9-202420      & 15.858$\pm$0.052 & 14.224$\pm$0.035 & 15.059$\pm$0.042 & 13.497$\pm$0.036 & 12.400$\pm$0.140  \\
GSC6209-735                         & 12.514$\pm$0.058 & 11.403$\pm$0.049 & 11.917$\pm$0.033 & 11.012$\pm$0.053 & 10.626$\pm$0.095 \\
GSC6794-156			     & 10.673$\pm$0.015 & 9.775$\pm$0.085 & 10.321$\pm$0.150 & 9.499$\pm$0.201 & 8.980$\pm$0.144\\
ScoPMS 17 			     & 15.571$\pm$0.366 & 13.833$\pm$0.095 & ...& ... & 12.058$\pm$0.086 \\
ROXs 47A				     & 15.381$\pm$0.098 & 13.611$\pm$0.095 & 14.510$\pm$0.119 & 12.835$\pm$0.092 & 11.615$\pm$0.051  \\
\enddata
\tablecomments{The B,V,g,r and i magnitudes taken from the latest APASS data release. Note that the photometry for ROXs 47A includes the wide tertiary component of the system, and so are not used in our fitting procedure below.}
\label{photometry_list}
\end{deluxetable*}

\begin{deluxetable*}{lccc}
\tabletypesize{\footnotesize}
\tablewidth{0.6\textwidth}
\tablecaption{2MASS Photometry}
\tablehead{
\colhead{Star}&\colhead{J}&\colhead{H}&\colhead{K}
}
\startdata
RXJ1550.0-2312  & 9.885$\pm$0.024 & 9.215$\pm$0.023 & 8.930$\pm$0.023\\
RXJ1601.9-2008	 & 8.350$\pm$0.020 & 7.808$\pm$0.026 & 7.672$\pm$0.020\\
USco J160517.9-202420  & 10.154$\pm$0.022 & 9.349$\pm$0.024 & 9.143$\pm$0.019\\
GSC6209-735    & 9.158$\pm$0.030 & 8.603$\pm$0.042 & 8.426$\pm$0.020 \\
GSC6794-156	 & 7.779$\pm$0.027 & 7.280$\pm$0.027 & 7.084$\pm$0.018\\
ScoPMS 17& 9.932$\pm$0.024 & 9.235$\pm$0.026 & 8.992$\pm$0.02 \\
ROXs 47A  & 9.245$\pm$0.024 & 8.351$\pm$0.031 & 7.929$\pm$0.061\\
\enddata
\tablecomments{The infrared are taken from the 2MASS catalog. Note that the photometry for ROXs 47A includes the wide tertiary component of the system, and so are not used in our fitting procedure below.}
\label{photometry_list_2mass}
\end{deluxetable*}

To reduce the dependence of the results of our age estimation on the characteristics of any one particular set of model isochrones, or at least to illuminate the model dependence, we use the Padova PARSEC 1.2S \citep{parsec12}, the Dartmouth PMS models \citep{dotter08}, and the BT-Settl Isochrones \citep{btsettl} to determine stellar properties of the binary systems in our Upper Scorpius sample. The Padova isochrones employ the  PHOENIX BT-Settl model atmospheres \citep{allard95,btsettl} for stars with effective temperatures cooler than 4000\,K,and ATLAS9 \citep{atlas9} model atmosphere for the hotter stars. These atmospheres are used for both the T$_{eff}$ to synthetic color transformations, and for establishing the relationship between temperature and the mean optical depth. This is then empirically adjusted based on colors of globular cluster stars,\citep{chen14_padova}.  Similarly, the Dartmouth models use the PHOENIX model atmospheres directly without further correction for all stars cooler that 10,000\,K, which spans the temperature range of all the primary stars included in this study. The BT-Settl isochrones are constructed by interpolating the \citet{btsettl} synthetic atmospheres over the \citet{baraffe98} and \citet{baraffe03} model grids, and are useable for stellar masses less than 1.4\,\Msun. 

There has been some issue with the color-magnitude relations for lower-mass stars (masses smaller than $\sim$1.2\,$M_{\odot}$): Models which closely reproduce observations in the near-IR and the redder optical colors often produce optical color significantly bluer than expected \citep{an08,dotter08}. The empirical corrections applied in the Padova isochrones was introduced to mitigate this issue for main-sequence dwarfs, and so we expect the Padova isochrones to more closely reproduce the photometry of our young binary systems. It is important to note however that the effectiveness of these corrections for PMS stars is not clear.

The photometry available to us for fitting varies between objects, for the most part,  B,V, g, r,  and i magnitudes from the APASS survey, which were of varying quality, and 2MASS near-IR magnitudes are available for all the target systems. The only absent photometry was the g and r band APASS magnitudes for ScoPMS 17. As mentioned previously, ROXs 47A is a hierarchical triple system \citep{barsony03}, of which we are examining the inner dynamical system. The third component of the system, which is of comparable brightness to the primary, is still relatively close to the primary, at  $0.79$'', which means that the 2MASS and APASS photometry are contaminated and unusable. Table \ref{photometry_list} tabulates all publicly available photometry for the stars in our sample, and for completeness, we include ROXs 47A in the table, though we do not use this photometry in the fitting procedure described above.  As mentioned above, we also incorporated our HST WFC3 wide band photometry and differential photometry obtained in 2012. For all the targets, we excluded the filters F225W, F275W, and F336W, because this wavelength range is highly sensitive to the activity of the PMS stars in question, and so cannot be reliably used to fit to models. We also empirically tested the effects of the F390W and F438W photometry on the fits for the M-type stars in the sample, but found that the inclusion or removal of these data did not significantly change the output results.

%%%%%%%%%%%%%%%%%%%%%%%%%%%

\subsection{Estimated Stellar Properties}

\begin{figure*}
\includegraphics[width=0.47\textwidth]{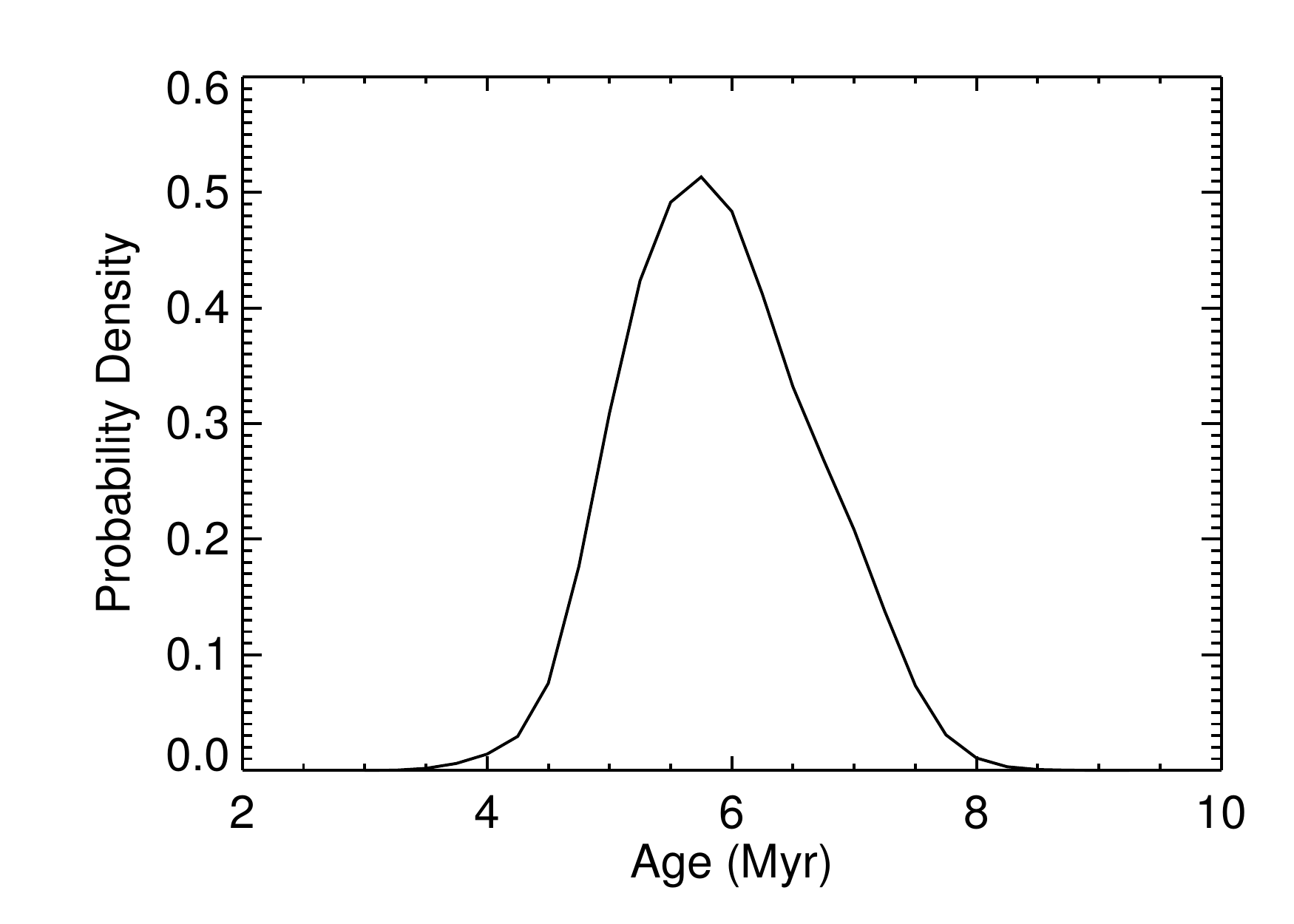}
\includegraphics[width=0.47\textwidth]{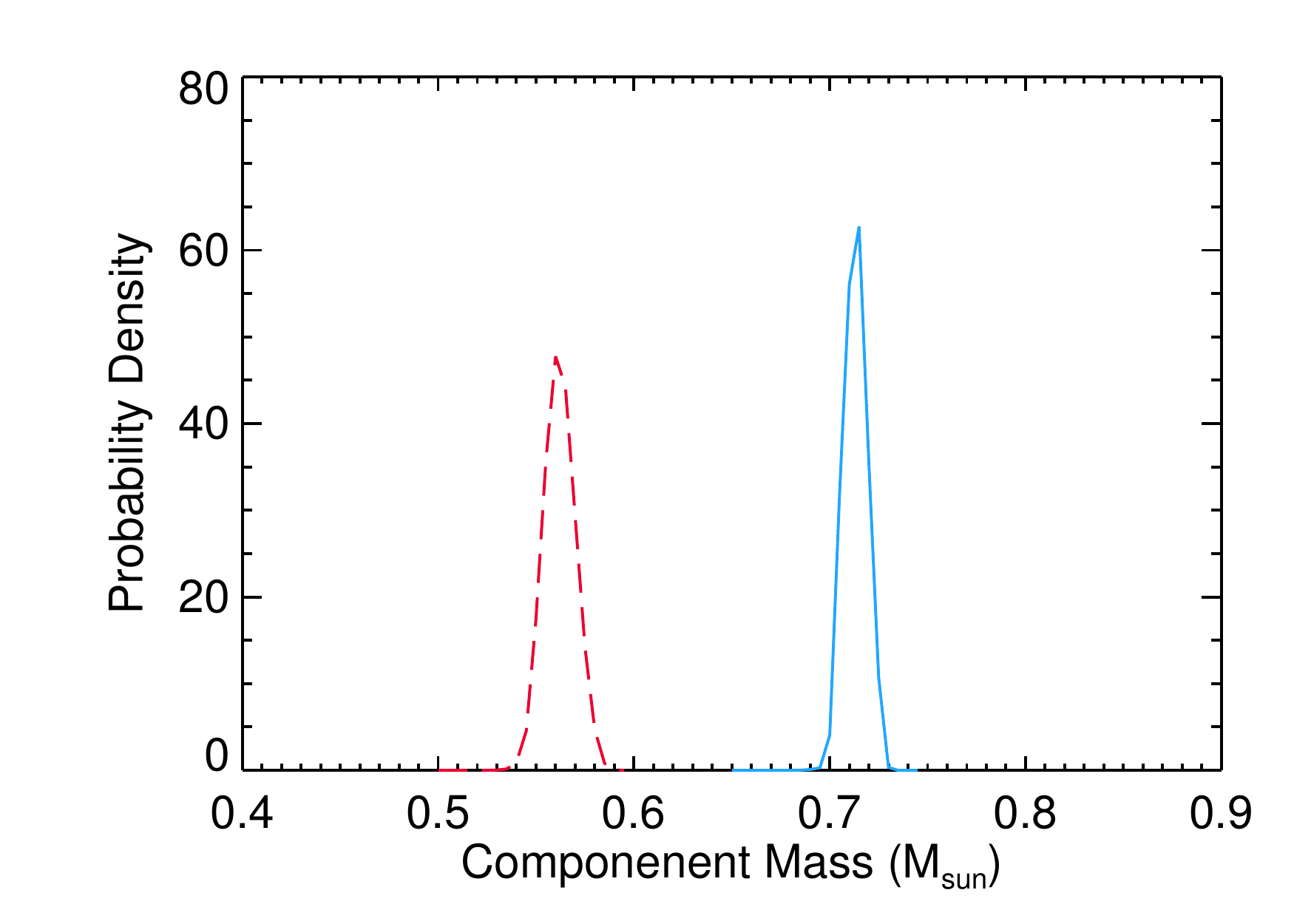}\\
\includegraphics[width=0.47\textwidth]{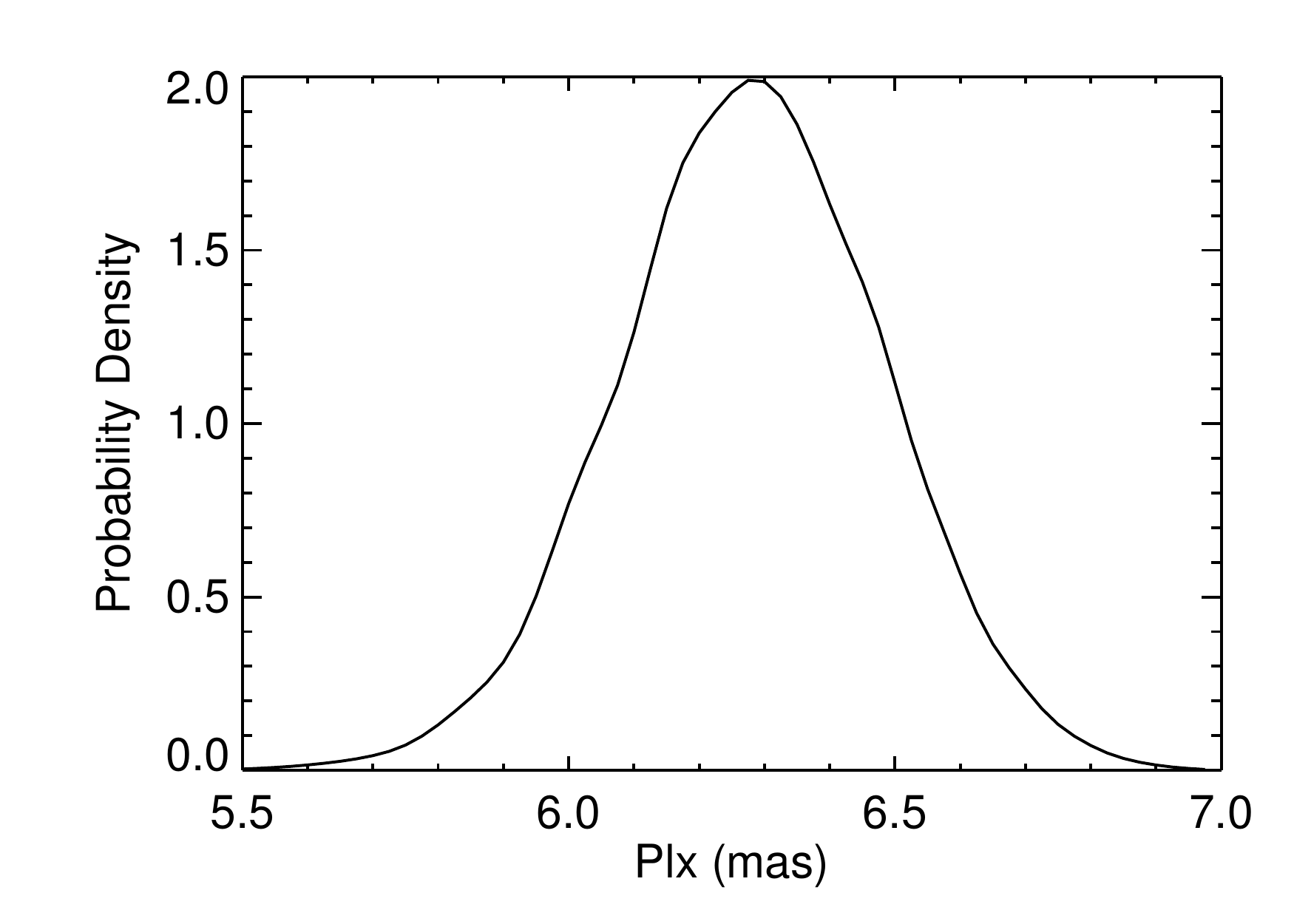}
\includegraphics[width=0.47\textwidth]{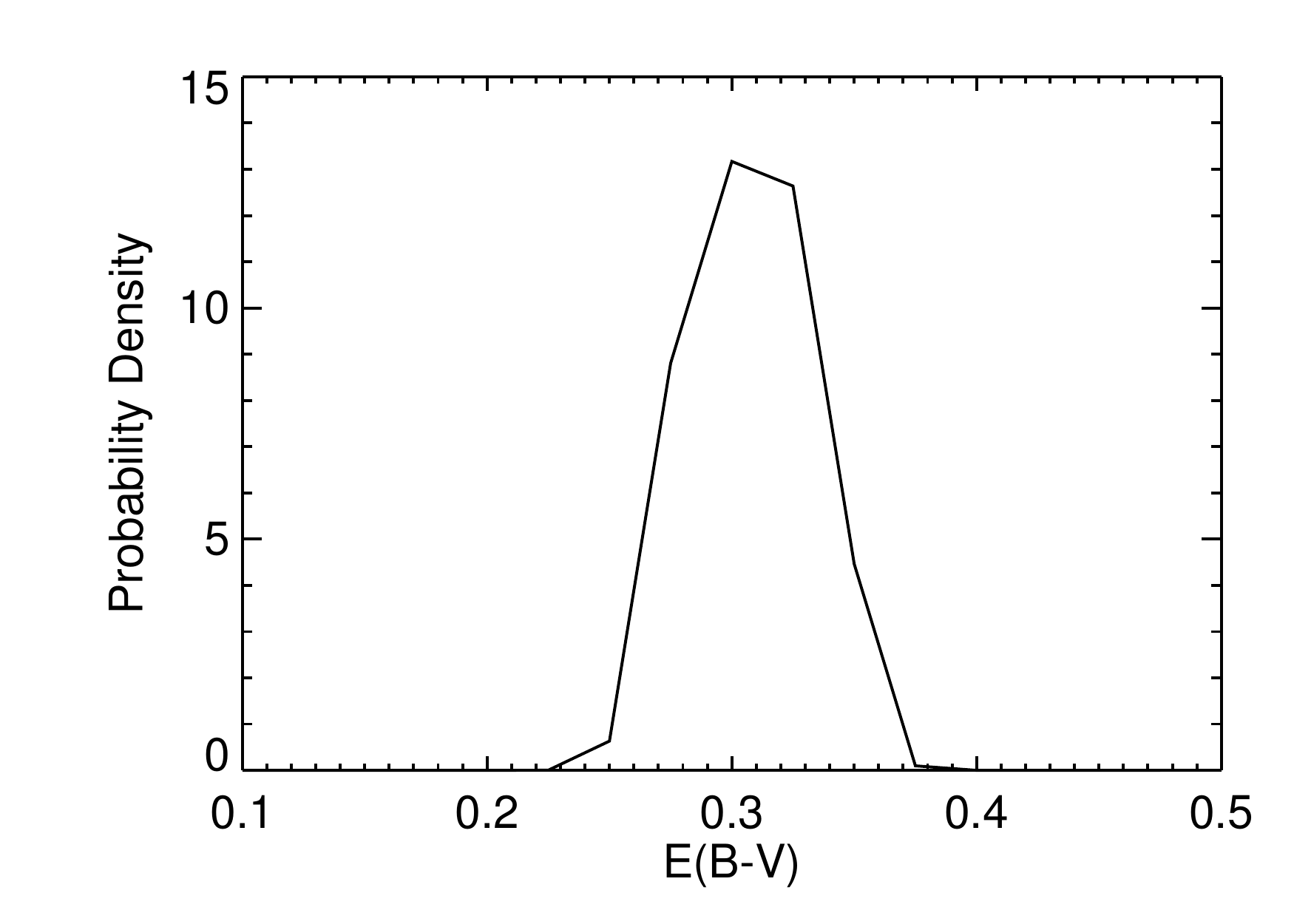}
\caption{Posterior distributions for each model parameter for the USco J160517.9-202420, fit with the Padova isochrones. Note that the two peaks in (b) are the primary (blue) and secondary (red) mass, which are placed on a single figure for ease of viewing, but are treated separately in the analysis. We show this set of posteriors as an demonstration that the output posteriors are Gaussian, the corresponding result for the rest of the binary sample and the other models, can be found in Table \ref{modrestab}.}
\label{P02-065_res}
\end{figure*}

Computation of the posterior probability for our Bayesian models yields a five dimensional space of probabilities, one dimension for each model parameter, which can be reduced to lower dimensionality by marginalizing over uncorrelated model parameters. We found none of our model parameters were strongly correlated, with the reddening parameter showing some correlation with the other parameters for some of the stars in our sample. This is expected given the possibility of degeneracy between distance, system mass and extinction as a function of age. We produce one dimensional probability densities for the parameters of our model, and then determine the intervals which contain the most likely values of each parameter. For all of the binary systems we have studied, a single most probable set of stellar parameters can be determined, with a corresponding uncertainty range which varied in size between systems. Figure \ref{P02-065_res} displays the probability of each model parameter for the star USco J160517.9-202420.

\begin{deluxetable*}{cccccccc}
\tabletypesize{\footnotesize}
\tablewidth{0pt}
\tablecaption{Bayesian Estimated Stellar Parameters}
\tablehead{
\colhead{Name}&\colhead{Model}&\colhead{Age} & \colhead{M$_p$}  & \colhead{M$_s$}& \colhead{$\pi$} & \colhead{$E(B-V)$} &\colhead{$\chi^2_\mathrm{r}$} \\
& & \colhead{(Myr)} &  \colhead{($M_\odot$)} & \colhead{($M_\odot$)} & \colhead{(mas)} & \colhead{(mag)}&
}
\startdata
GSC6794-156     & P & 10.5$^{+1.3}_{-1.7}$      & 1.51$\pm$0.1        & 1.41$\pm$0.1        & 7.3$\pm$0.2          & 0.40$\pm$0.12        & 6.4  \\
                & D & 10.2$^{+1.6}_{-1.0}$      & 1.46$\pm$0.2        & 1.38$\pm$0.2        & 7.3$\pm$0.2          & 0.37$\pm$0.10       & 7.3  \\
RXJ1601.9-2008  & P & 12.8$\pm 1.5$           & 1.46$\pm$0.05        & 0.82$\pm$0.04        & 6.8$\pm$0.2          & 0.53$\pm$0.10        & 6.8   \\
                & D & 11.5$\pm 2.2$           & 1.45$\pm$0.02        & 0.77$\pm$0.04        & 6.8$\pm$0.3          & 0.43$\pm$0.07        & 7.3   \\
GSC6209-735     & P & 18.8$\pm 5.6$           & 1.14$\pm$0.09        & 0.45$\pm$0.02        & 8.0$\pm$0.7          & 0.41$\pm$0.09        & 5.6   \\
                & D & 23.8$^{+5.8}_{-6.8}$      & 1.23$\pm$0.15        & 0.27$\pm$0.03        & 8.1$\pm$1.1          & 0.57$\pm$0.09        & 8.6   \\
                & B & 27.4$\pm3.1 $           & 1.16$\pm$0.04        & 0.25$\pm$0.03        & 7.7$\pm$0.2          & 0.41$\pm$0.10        & 5.4   \\
USco J1605...   & P & 5.9$^{+0.9}_{-1.2}$       & 0.71$\pm$0.01        & 0.56$\pm$0.01        & 6.3$\pm$0.3          & 0.31$\pm$0.04        & 1.7   \\
                & D & 5.4$\pm 0.8$            & 0.61$\pm$0.03        & 0.45$\pm$0.02        & 6.6$\pm$0.3          & 0.29$\pm$0.04        & 1.7   \\
                & B & 6.5$\pm 0.7$            & 0.65$\pm$0.03        & 0.49$\pm$0.03        & 6.5$\pm$0.3          & 0.29$\pm$0.02        & 1.3   \\
ScoPMS 17       & P & 7.1$\pm 0.9$            & 0.71$\pm$0.01        & 0.45$\pm$0.01        & 7.5$\pm$0.1          & 0.21$\pm$0.03        & 2.3   \\
                & D & 6.5$\pm 0.7$            & 0.59$\pm$0.03        & 0.35$\pm$0.02        & 8.0$\pm$0.2          & 0.21$\pm$0.02        & 1.6   \\
                & B & 7.4$\pm 0.7$            & 0.59$\pm$0.03        & 0.36$\pm$0.02        & 8.0$\pm$0.2          & 0.19$\pm$0.03        & 1.4   \\
RXJ1550.0-2312  & P & 18.3$\pm 1.9$           & 0.70$\pm$0.02        & 0.43$\pm$0.02        & 10.6$\pm$0.3         & 0.26$\pm$0.06        & 3.1   \\
                & D & 14.0$\pm 2.7$           & 0.54$\pm$0.08        & 0.31$\pm$0.05        & 11.7$\pm$0.7         & 0.28$\pm$0.06        & 10.6   \\
                & B & 14.4$\pm 2.7$           & 0.47$\pm$0.08        & 0.27$\pm$0.05        & 12.1$\pm$0.7         & 0.26$\pm$0.05        & 8.6   \\
ROXs 47A        & P & 3.8$\pm 1.6$            & 0.73$\pm$0.03        & 0.68$\pm$0.03        & 7.2$\pm$0.1          & 0.53$\pm$0.08        & 5.3   \\
                & D & 3.5$\pm 2.3$            & 0.61$\pm$0.06        & 0.56$\pm$0.06        & 7.6$\pm$0.3          & 0.45$\pm$0.12        & 10.5  \\
                & B & 5.2$\pm 1.6$            & 0.85$\pm$0.11        & 0.78$\pm$0.08        & 6.8$\pm$0.3          & 0.48$\pm$0.08        & 7.1   \\
\enddata
\tablecomments{The models Padova (P), Dartmouth (D), and BT-Settl (B) refer to the \citet{padova02}, \citet{dotter08} and \citet{btsettl} model grids respectively. The final column lists the model best fit reduced $\chi^2$ value.}
\label{modrestab}
\end{deluxetable*}

Given these output distributions, we then calculate 1-$\sigma$ Bayesian ``credible" intervals for each parameter, which can be found in Table \ref{modrestab}, where we have scaled the uncertainties by the reduced chi-squared of the model fits. Note that the BT-Settl isochrones were only applied to the binary systems in our sample where the component masses were expected to be less than 1\,\Msun. We find that the output most likely system parameters agree with the estimates obtained from the conventional SED-Fitting and HR-diagram positions. Below, we individually summarize the results for each binary system, detailing the clarity of each fit.

\subsubsection{GSC 6794-156}
Both the Padova and Dartmouth model fits produce a system age of $\sim$10.5\,Myr for GSC 6794-156, which is consistent with the recent \citet{pecaut12} age estimation for Upper Scorpius. In the other four parameters the models agree closely. The best fit system parallax for both models is 7.3$\pm$0.2\,mas, which is consistent with the mean distance and distance spread in Upper Scorpius. We find that the Padova and Dartmouth models produce a most likely reddening parameter of $E(B-V)\simeq0.43$\,mag. Given both the $\Delta J$ and $\Delta K$ values from the Keck NIRC2 aperture masking (see Table \ref{orbital_points_nrm}), we can estimate the expected extinction for this system using standard tables of template photometry for young systems, with some uncertainty produced by the unclear spectral-type of the primary. The tables of \citet{pecaut13} give an intrinsic J-K color of  of 0.47\,mag for the approximately G6 primary of system this system. The observed color, corrected for the presence of the companion using the aperture masking contrasts is 0.617\,mag, which then yields a value of $E(B-V) \sim 0.2-0.4$\,mag, which i consistent with our model fit value.

\subsubsection{RXJ1601.9-2008} 
Both the Padova and Dartmouth models produce consistent ages of $\sim$12\,Myr for the RXJ1601.9-2008 binary system, and we note that as with the other G-type binary system in our sample (GSC 6794-156) the Padova age estimate is slightly older than that of the Dartmouth model. The other four system parameters agree closely between the models, and we note that both RXJ1601.9-2008 and GSC 6794-156 have very similar primary component masses. This is expected given the similar spectral types for these two systems (G5 and G6 respectively). 

\subsubsection{GSC 6209-735}
The visual orbit for GSC 6209-735 is the least well-constrained in our sample due to the large contrast ratio between the primary and secondary ($\Delta$H$=$3.05\,mag) and the small angular separation. Fortunately, GSC 6209-735 has been known to be a single lined spectroscopic binary for some time, and the single component radial velocity orbit has been previously determined \citep{guenther07}. We included the single lined orbit information in the model fit as a second mass observable as described above and were able to produce posteriors for the stellar parameters that are relatively well-constrained. We find that the component masses and the system parallax determined for the three models agree within the uncertainties, with the system parallax of 7.9$\pm$0.2\,mas being consistent with Upper Scorpius membership. The Dartmouth and BT-Settl models produce most probable ages of $\sim$23-26\,Myr, while the Padova isochrones produce and age of $\sim$19\,Myr. The possible age solutions for this binary system are significantly older than the other PMS binary systems in our sample, and places GSC 6209-735 as a potential member of the older Upper Centaurus Lupus subgroup of Sco-Cen. The estimated  parallax, and location of GSC 6209-735 near the centre of Upper Scorpius are consistent with both Upper Scorpius and UCL. We also note that GSC 6209-735 is the only K-type star in our sample and shows the largest age discrepancy of the binary systems we have monitored.

\subsubsection{USco J160517.9-202420}
We find that the best fit model parameters for USco J160517.9-202420 indicate that it is a young binary system of age $\sim$6\,Myr, with the three model fits producing age and parallax estimates that agree very closely. There is some difference in the best fit mass between the models: The Padova model produces primary and secondary masses which are significantly larger than the corresponding Dartmouth and BT-Settl model fits.  The study in which this star was identified as a Sco-Cen member estimates $E(B-V) = 0.3$ \citep{preibisch02}, and estimation using spectral type (M3) and $J-K$ color yields $E(B-V)=0.2\pm0.1$ using the intrinsic color tables of  \citet{pecaut13}. Both these estimations are consistent with our estimates. 

\subsubsection{ScoPMS 17}
 All three models produce an age consistent with $\sim$7\,Myr for ScoPMS 17, however, as with USco J160517.9-202420, the Padova models produce significantly larger masses for both the primary and secondary components of the binary system. Estimation from the color tables using the spectral type (M1) and the $J-K$ color gives a value of  $E(B-V)=0.1\pm0.1$ \citep{pecaut13}, which is consistent with out model fits.

\subsubsection{RXJ 1550.0-2312}
From the original orbital solution for RXJ1550.0-2312, there was clear evidence that the system is significantly closer than the median Upper-Scorpius parallax of $\sim$6.9\,mas (See Table \ref{orb_params}). Upon applying the Bayesian fitting method described above, we found a peak in the system parallax PDF beyond 10\,mas, and so we removed the input prior system parallax of 7.5$\pm$1.6\,mas and refit the data for the three models. The model age fits for RXJ 1550.0-2312 are significantly older than those of the other M-type Upper Scorpius binary systems we have studied ($>$14\,Myr), and we note that the Padova age estimate is older still than that of the other models. As with the other M-type systems in this study,  the Padova mass estimates are significantly larger than that of the Dartmouth or BT-Settl models. The distance and age estimates for RXJ 1550.0-2312 mean that it is unlikely to be a member of Upper Scorpius, however the detection of Li absorption in the spectra \citet{preibisch01} confirms that is must definitely be young ($\lesssim$20\,Myr).  RXJ1550.0-2312 sits in the region of sky ($l,b$ = $347.8^\circ, 23.7^\circ$) traditionally considered the border between Upper Scorpius and the older ($\sim$16\,Myr) Upper-Centaurus-Lupus (UCL) subgroup of Sco-Cen \citep{zeeuw99}, and has proper motions of (-20.1$\pm$1.0, -25.8$\pm$1.8)\,mas  \citep{ucac4}, which are only marginally consistent with UCL. If we take these proper motions and the distance estimate from our model fitting, and apply the membership probability method outlined in \citet{myfirstpaper} and \citet{wifes1_2015} using the UCL mean space velocity from \citet{chen11} of $(U,V,W)=(-5.1,-19.7,-4.6)$\,km/s, we find that RXJ 1550.0-2312 has a 57\% probability of membership in UCL.  This indicates that RXJ 1550.0-2312 may be a peripheral member of UCL, or a young foreground star.

\subsubsection{ROXs 47A}
We expected ROXs 47A to be the most difficult M-type system for the models to accurately reproduce, given the very young age and disk presence.
The Bayesian fitting procedure produced a very young and highly reddened fit for ROXs 47A, with an estimated age $<4$\,Myr for the Padova and Dartmouth model. The estimated parallaxes for both the Padova and Dartmouth models (7.8\,mas and 7.7\,mas respectively) are consistent with ROXs 47A being a members of the very young $\rho$-Ophiuchus star forming region, which is located at $(\alpha,\delta) = (16^h 28^m, -24^\circ33'')$ and a distance of $\sim$130\,pc, i.e., slightly closer than the Upper-Scorpius subgroup. The BT-Settl models produce significantly different fits to the data, with very large component masses of 0.85\,\Msun and 0.78\,\Msun respectively, and a parallax of 6.9$\pm$0.2\,mas. This parallax is inconsistent with the distance to Ophiuchus,  and the estimated age of 5.2$\pm$0.5\,Myr is significantly older than the mean age of 2.1\,Myr of the Ophiuchus PMS stars \citep{mcclure2010}. All three fits produce extinction values consistent with extinction value of  E(B-V)=0.52\,mag estimated by \citet{mcclure2010}. 

\section{Evaluation of the PMS Models}

With the results of the Bayesian model fitting procedure described above, it is possible to make informed statements about the behavior of the models for predicting stellar properties of young stars of different mass and spectral type, and the difference in outcome between the varieties of models for evaluating a single binary system. 

Firstly, we see that for the M-type binary systems, the fitted component masses are highly dependent on the choice of model, with the Padova models producing significantly larger component masses than the Dartmouth or BT-Settl models, while the other four parameters are in general agreement. This is consistent with recent observations of UScoCTIO 5, an M4.5 eclipsing binary in Upper Scorpius. \citet{kraususcoctio5} found the model masses produced for UScoCTIO 5 were overestimated by the Padova models and underestimated by the Dartmouth and Baraffe models \citep{baraffe15}. In contrast to this trend, the fitted parameters for the G-type stars in our sample agree very closely between the Padova and Dartmouth models, which predict very similar component masses and system parallaxes.

A number of the binary systems we have characterized appear to be younger than 10\,Myr according to the models, while other binary systems in our sample are of age $>10$\,Myr. The \citet{pecaut12} study, which used photometry for B, A, F and G-type stars, estimated the age of the Upper Scorpius subgroup to be 11$\pm$2\,Myr, which is significantly different to the age determinations for the younger stars in our work, and the age estimations in previous work \citep{preibisch02,geus92}. 

Most importantly, we find that across the three models, there is a significant dichotomy in the estimated ages, with the G-type members appearing older ($\sim$12\,Myr) compared to the M-type members ($\sim$7\,Myr). This trend was broadly mirrored in the conventional HR-Diagram position ages from the binary SED fits. In this comparison we exclude RX J1550.0-2312 because it has distance and age estimates consistent with the older Upper Centaurus Lupus subgroup of Sco-Cen, and GSC 6209-735 which appears discrepantly too old to be a member of Upper Scorpius and is the only K-type stars in our study, making attribution of the age discrepancy to either the models or membership difficult. We have also removed ROXs 47A from this comparison due to it's expected youth as a member of Ophiuchus star forming region. The apparent age difference between the remaining four G and M-type binary systems we see here is consistent with previous HR-diagram ages for different spectral type populations in Upper Scorpius: \citet{preibisch02} found that the M-type members in Upper Scorpius had a mean age between 3-5\,Myr based on HR-diagram estimation, and more recent work involving a new sample of Upper Scorpius PMS M-type members and modern isochrones found similar results, with the later M-type members showing an overall younger age than K-type members \citep{wifes1_2015}.

The most recent main-sequence A-type, PMS F-type, and PMS G-type member age estimates are 9$\pm$2\,Myr, 13$\pm$1\,Myr,  and 10$\pm$3\,Myr respectively \citep{pecaut12}, all of which broadly agree with our Bayesian age estimate for the G-type binaries in our orbit monitoring sample. We thus conclude that the evolutionary models for the M-type stars, which we find to produce ages of $\sim6-7$\,Myr, do not adequately reproduce the descent towards the main-sequence along the Hayashi track \citep{hayashitrack}. \citet{karnath13} also found that for two PMS spectroscopic binaries in NGC2264, model ages for the  M-type secondaries were consistently consistently younger than those for the G/K-type primaries. This age discrepancy is consistent with HR-diagram ages of high-mass members \citep{pecaut12} compared to M-type members \citep{preibisch01,wifes1_2015} in Upper Scorpius, and is also consistent with the age discrepancy estimations from \citet{naylor09} and  \citet{bell13}. In particular, the age discrepancy found here is equivalent to an under-prediction of the luminosity of a PMS M-type star of a given mass at a given PMS age, with the typical luminosity under prediction of 0.08 to 0.15\,dex depending on the particular model and the stellar mass. 

Given that distance is a free parameter in our analysis, it is difficult to disentangle whether the age discrepancy is caused by a mis-calibration in the temperature scales of the models or the luminosity evolution of the star in time.  \citep{hillenbrand04} found that adopting a warmer than dwarf temperature scale for pre-main-sequence binaries could improve the match between observations and the models. In light of this, and the recent studies of the Upper Scorpius M-type eclipsing binaries, in particular UScoCTIO5 \citep{kraususcoctio5} and other systems \citep{david15}, which indicate that in the case of a luminosity independent test, the model luminosities are generally consistent with measurements, while the temperature predictions are incorrect, we recast the discrepancies in age that we observe in terms of effective temperature. Assuming that varying the system distance can produce agreement in both luminosity and system masses for a given system, this luminosity discrepancy is equivalent to a 100-300\,K overestimation of the effective temperature of the binary components depending on stellar mass. This clear in Figure \ref{gmhrd}, where shifting the isochrone grid to cooler temperatures give older ages for the M-type binary system components. 

\begin{figure}
\includegraphics[width=0.5\textwidth]{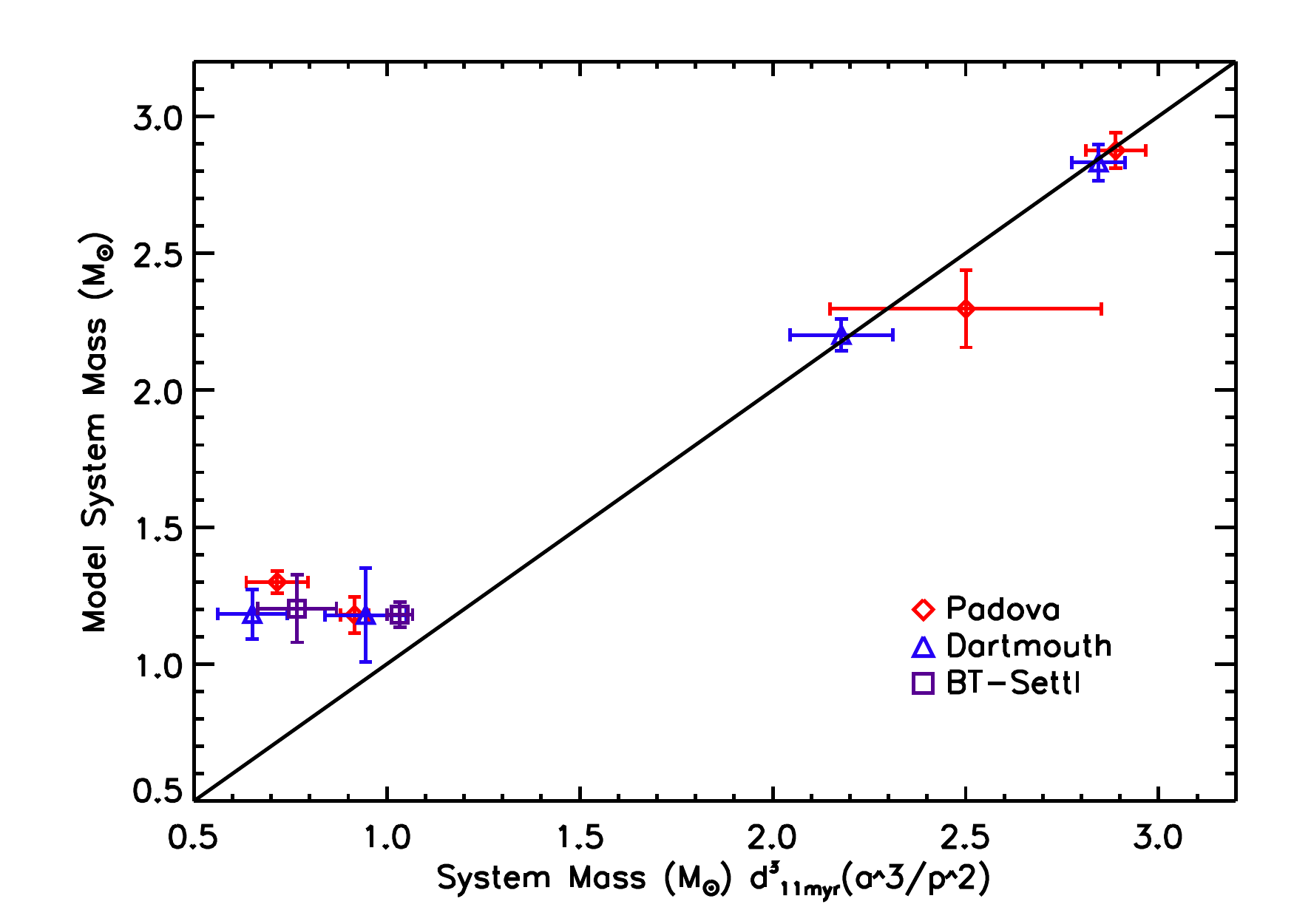}
\caption{Measured system mass and model system mass at a fixed age of 11\,Myr for the two G-type and two M-type stars in our sample that are clear USco members. The measured system mass in the x-axis is derived from the orbital semimajor axis and period but is dependent of the model distance at the fixed age of 11\,Myr ($d^3_{11Myr}a^3/p^2$). The model system masses for the M-type stars are significantly overestimated compared to the measured masses, while the G-type stars show very close agreement between the model fits and the data.}
\label{sysmass_comp}
\end{figure}

To visualize the discrepancy between the models and the data in terms of the component masses, we fix the system age in our Bayesian fitting procedure to 11\,Myr, and allow the component masses, parallax and reddening parameter to be varied. We include in this comparison the two G-type stars GSC 6794-156 and RXJ 1601.9-2008, and the two clear M-type members of  USco, J160517.9-202420 and ScoPMS 17. Figure \ref{sysmass_comp} displays the fitted model system mass at 11\,Myr compared to the measured system mass at the best-fit 11\,Myr distance for each system. For the G-type stars, we find, as expected that the model system masses agree closely with the measured system masses. The model system masses for the M-type binary systems are systematically larger, on the order of 0.2-0.4\,M$_\odot$, than the measured values.

The primary limitation in pinpointing the exact mis-calibration in the sub-solar regime of the pre-main-sequence models is the absence of a precise measure of the binary system distances. The best available measure at this time is association membership ($\pm$15\,pc), and does not allow disentanglement of the degeneracies between luminosity, dynamical mass, and distance. The upcoming  availability of high-precision parallaxes from the GAIA mission, combined with the orbital precision obtained for these systems, will break the current degeneracies in what could be causing the M-dwarf discrepancies.

\section{Summary}

We have presented astrometric orbits and HST WFC3 photometry of seven G,K and M-type binary systems in the young ($\sim$10\,Myr) Upper Scorpius subgroup of Sco-Cen. Using the orbital parameters and multi-band photometry we have determined estimated system parameters based on various model isochrones using a Bayesian fitting technique. The model stellar properties for the seven binary systems allow us to conclude that:

\begin{itemize}
\item The model isochronal ages derived from fitting to the Padova, Dartmouth and BT-Settl isochrones for the G-type binary systems is $\sim$11.5\,Myr, which is closely consistent with the latest HR-diagram age for Upper Scorpius \citep{pecaut12}. 
\\
\item The mass predictions for the M-type binary systems differ between the models, with the Padova models predicting significantly larger binary component masses then the Dartmouth and BT-Settl isochrones.
\\
\item For the M-type binary systems, the isochronal ages are $\sim$7\,Myr, which is significantly younger than expected (11\,Myr) and indicates calibration issues in the models for the M-type regime. This age discrepancy is equivalent to a Luminosity under-prediction of 0.08-0.15\,dex, or an effective temperature over-prediction of 100-300\,K. This suggests both the possibility of an uncertain temperature scale or further calibration issues in the mass-radius relation.

\end{itemize}

\acknowledgements
Some of the work presented here is based on observations made with the NASA/ESA Hubble Space Telescope, obtained from the data archive at the Space Telescope Science Institute. STScI is operated by the Association of Universities for Research in Astronomy, Inc. under NASA contract NAS 5-26555. Some of the data presented herein were obtained at the W.M. Keck Observatory, which is operated as a scientific partnership among the California Institute of Technology, the University of California and the National Aeronautics and Space Administration. The Observatory was made possible by the generous financial support of the W.M. Keck Foundation.The authors wish to recognize and acknowledge the very significant cultural role and reverence that the summit of Mauna Kea has always had within the indigenous Hawaiian community.  We are most fortunate to have the opportunity to conduct observations from this mountain.

%%REFERENCES%%%%%%%%%%%%%%
\bibliographystyle{apj}
\bibliography{ms}
%%%%%%%%%%%%%%%%%%%%%%%%%

\end{document}